\documentclass[CRPHYS,Unicode,manuscript]{cedram}
\usepackage{graphicx}
\usepackage[utf8]{inputenc}
\usepackage{dsfont}
\usepackage{empheq,etoolbox}
\usepackage{ amssymb }

\usepackage{chngcntr}
\usepackage{array}

\newcommand{\FT}[2]{\mathrm{FT}\left[#1\right]\left(#2\right)}
\newcommand{\FTinv}[2]{\mathrm{FT}^{-1}\left[#1\right]\left(#2\right)}
\newcommand{\iup}{_{\mathrm{up}}}
\newcommand{\ido}{_{\mathrm{down}}}
\newcommand{\DM}{_{\mathrm{DM}}}

\hypersetup{
    colorlinks=true,
    linkcolor=blue,
    citecolor=red,
    urlcolor=teal}

\title{Imaging exoplanets with coronagraphic instruments}

\alttitle{Imager des exoplanètes grâce aux instruments coronographiques}

\author{\firstname{Raphaël} \lastname{Galicher}}
\address{LESIA, Observatoire de Paris, Université PSL, CNRS, Université Paris Cité, Sorbonne Université, 5 place Jules Janssen, 92195 Meudon, France}
\email[]{raphael.galicher@obspm.fr}

\author{\firstname{Johan} \lastname{Mazoyer}\CDRorcid{0000-0002-9133-3091}}
\addressSameAs{1}{LESIA, Observatoire de Paris, Université PSL, Sorbonne Université, Université Paris Cité, CNRS, 5 place Jules Janssen, 92195 Meudon, France}
\email[]{johan.mazoyer@obspm.fr}


\altkeywords{Exoplanètes, Instrumentation astronomique, Coronographie, Imagerie Haut-contraste, Haute résolution angulaire}

\begin{abstract} 
Exoplanetary science is a very active field of astronomy nowadays, with questions still opened such as how planetary systems form and evolve (occurrence, process), why such a diversity of exoplanets is observed (mass, radius, orbital parameters, temperature, composition), and what are the interactions between planets, circumstellar disk and their host star. Several complementary methods are used for the detection of exoplanets. Among these, imaging aims at the direct detection of the light reflected, scattered or emitted by exoplanets and circumstellar disks. This allows their spectral and polarimetric characterization. Such imaging remains challenging because of the large luminosity ratio ($10^4$-$10^{10}$) and the small angular separation (fraction of an arcsecond) between the star and its environment. Over the past two decades, numerous techniques, including coronagraphy, have been developed to make exoplanet imaging a reality.

This paper gives a broad overview of the subsystems that make up a coronagraphic instrument for imaging exoplanetary systems. It is especially intended for non-specialists or newcomers in the field. We explain the principle of coronagraphy and propose a formalism to understand their behavior. We discuss the impact of wavefront aberrations on the performance of coronagraphs and how they induce stellar speckles in the scientific image. Finally, we present instrumental and signal processing techniques used for on-sky minimization or a posteriori calibration of these speckles in order to improve the performance of coronagraphs.

\end{abstract}

\begin{altabstract} 
L'exoplanétologie est un domaine très actif de l'astronomie moderne avec des questions encore ouvertes: comment les systèmes planétaires se forment-ils et évoluent-ils ; pourquoi une telle diversité d'exoplanètes est-elle observée (masse, rayon, paramètres orbitaux, température, composition) ; quelles sont les interactions entre les planètes, les disques circumstellaires et leur étoile hôte ? Plusieurs méthodes complémentaires sont utilisées pour la détection d'exoplanètes. Parmi celles-ci, l'imagerie permet la détection directe de la lumière réfléchie, diffusée ou émise par les exoplanètes et les disques circumstellaires. Ceci permet une caractérisation spectrale et polarimétrique. Obtenir une image d'exoplanète n'est cependant pas simple en raison du grand rapport de luminosité ($10^4$-$10^{10}$) et de la faible séparation angulaire (fraction de seconde d'angle) entre l'étoile et son environnement. Depuis deux décennies, de nombreuses techniques, dont la coronographie, ont été développées pour faire de l'imagerie des exoplanètes une réalité.

Cet article donne un large aperçu des sous-systèmes d'un instrument coronographique. Il a été écrit en particulier pour les non-spécialistes ou les nouveaux venus dans le domaine. Nous décrivons le fonctionnement de la coronographie et en proposons un formalisme mathématique. Nous expliquons la formation des tavelures stellaires et l'impact des aberrations de la surface d'onde sur les performances du coronographe. Nous présentons enfin les techniques instrumentales et de traitement du signal utilisées pour améliorer les performances des coronographes en minimisant activement ou en étalonnant a posteriori ces tavelures.

\end{altabstract}

\begin{document}

\maketitle

\section{The challenge of exoplanet imaging}

The goal of direct imaging is to obtain images of the circumstellar environments of stars: exoplanets, debris and protoplanetary disks. In this review, and more generally in our field, {\it image} refers to broadband images or integral field spectrometer data (an observational technique that provides spectral information over a 2D field of view). Imaging allows direct access to the exact position of the source around the star as well as a measurement of the light reflected or emitted by the object. For exoplanets, this allows for the determination of both orbital parameters and physico-chemical properties of its atmosphere. Direct imaging is however challenging because of the significant luminosity ratio (from $10^{-4}$ to $10^{-10}$) and small projected angular separations (from a few hundredths of an arcsecond to a few arcseconds for the closest stars) between a planet and its host star. For these reasons, less than~$1\,\%$ of the~$\sim5000$ exoplanets discovered to date have been directly imaged.

\begin{figure}[!ht]
    \centering
    \includegraphics[width=\textwidth]{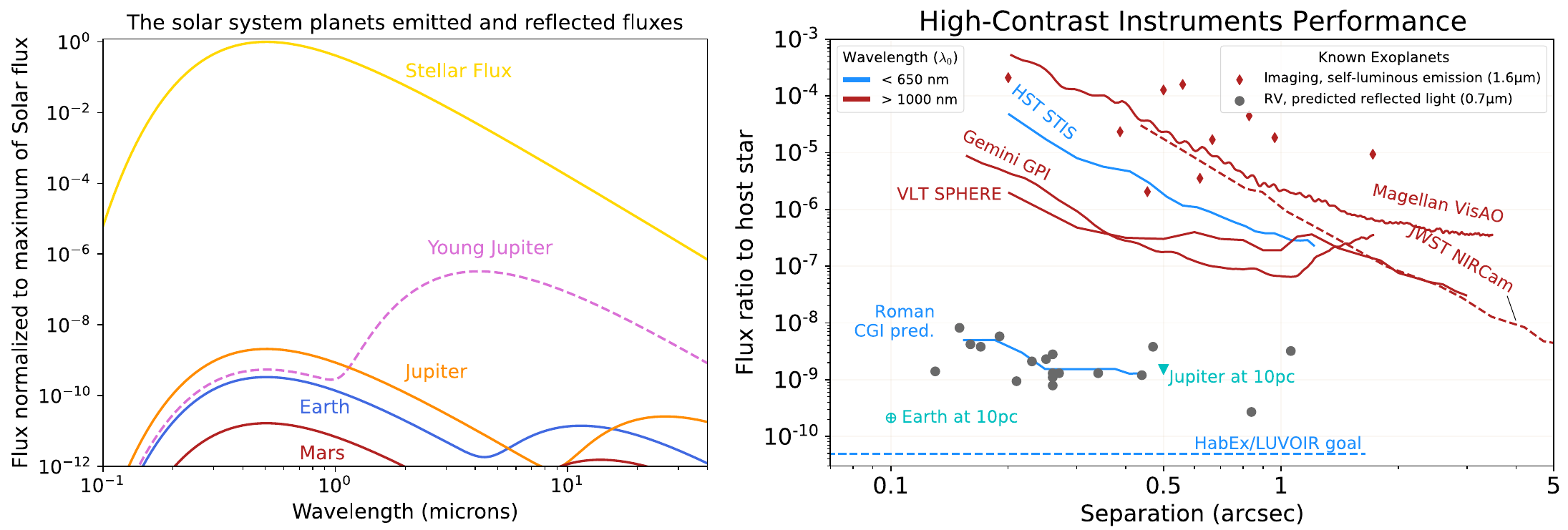}
    \caption{\it \textbf{Left:} Reflected (in the visible) and emitted (in the near-IR) fluxes for Solar System planets, normalized to the maximum of the Sun flux.
    Temperatures and albedos from \cite{encrenaz2013_SolarSystem}. The dashed curve plots a young Jupiter flux using physical values of the 51~Eridani system \cite{macintosh2015_DiscoverySpectroscopyYoung}. \textbf{Right:} $5\,\sigma$ detection limits of ground- and space-based instruments. Red diamonds represent few of the currently imaged exoplanets, with their known planet to star flux ratio in~H-band. Grey circles show known exoplanets detected by the RV method, for which the reflected light flux ratio is predicted in the visible. Adapted from \cite{baileywebsite_DIfluxratioplotFluxRatio}.}
    \label{fig:required_contrast}
\end{figure}

The plot on the left in Figure~\ref{fig:required_contrast} shows the emitted (infrared) and reflected (visible) fluxes of the planets in the solar system, normalized to the maximum of the stellar flux.
This shows that the luminosity ratio between the exoplanet and its host star is more favorable for imaging in the infrared, where the exoplanet emission peaks ($\sim10\,\mu$m for Solar system planets), but where the angular resolution, which degrades linearly with the wavelength, is poor. In the visible, the angular resolution is better but the flux ratio is very large. A trade-off can be found with young, massive Jupiter-like planets at~$\sim1\,\mu$m (dashed line). Such a planet is warmer than the Solar system planets and the maximum of its emission is at~$1-3\,\mu$m for which the angular resolution of the telescope is $3-10$ times smaller than at~$10\,\mu$m. Hence, current exoplanet imaging instruments have been optimized for near-infrared observations~(Y to L band) to detect these warm Jupiters.

Most of the current 8m-class ground-based telescopes host or have recently hosted high-contrast instruments that include a stellar coronagraph. The pioneering instruments were developed in the late~1990s on the first generation of Adaptive optics~(AO) imagers:  VLT/NACO \cite[first light~2001, decommissioned~2019]{lenzen2003_NAOSCONICAFirstOnsky, rousset2003_NAOSFirstAO}, Keck/NIRC2\cite[first light~2001, still active]{McLean2003_instru_Keck,Wizinowich2006_keck} and Gemini South/NICI \cite[first light~2007, decommissioned~2014]{toomey2003_InfraredCoronagraphicImager}  and Gemini North/NIRI \cite{Hodapp03,Herriot00}. The lessons learned from these instruments lead to a second generation of instruments a decade later. As the first generation of instruments consisted mainly of multi-purpose~AO assisted~IR imagers with, among other modes, one coronagraphic channel, the second generation was designed much more specifically for high-contrast imaging of exoplanets, combining extreme AO systems with recent advances in coronagraphy: VLT/SPHERE \cite[first light~2014, still active]{beuzit2019_SPHEREExoplanetImager} , Gemini South/GPI \cite[first light~2014, currently being upgraded off telescope since~2020]{macintosh2008_GeminiPlanetImager}, MagAO/Clio2 \cite{sivanandam2006_Clio35Micron, close2010_MagellanTelescopeAdaptive} and Subaru/SCExAO \cite[first light~2017, still active]{jovanovic2015_SubaruCoronagraphicExtreme}. 

In space, the extreme sensitivity and optical stability of the Hubble space telescope~(HST) makes it an ideal facility for high-contrast imaging. Three~HST instruments include a coronagraphic channel, NICMOS (operational from~1997 to~1999 and from~2002 until~2008 \cite{schneider1998_ExplorationEnvironmentsNearby}),~ACS (installed in~2002 and observing in the visible, still active but the high-resolution channel which included the coronagraph was permanently disabled in~2007 \cite{krist2003_AdvancedCameraSurveys}) and~STIS (operational from~1997 to~2004 and from~2009 until now, observing in the visible \cite{grady2003_CoronagraphicImagingHubble}). The~HST instruments, designed long before the first exoplanet was imaged, are multi-purpose, with rudimentary coronagraphs that are not optimized for exoplanet imaging. Finally, the recently launched JWST includes two instruments with coronagraphic modes: NIRCam \cite{beichman2012_ScienceOpportunitiesNearIRa} in the near-infrared and MIRI in the mid-infrared \cite{boccaletti2015_MidInfraredInstrumentJames}.

Some of the most iconic discoveries of recent years include 51~Eri~b \cite{macintosh2015_DiscoverySpectroscopyYoung}, HD~95086~b \cite{rameau2013_ConfirmationPlanetHD}, HR~8799~bcde \cite{marois2008_DirectImagingMultiple, marois2010_ImagesFourthPlanet}, HIP~65426~b \cite{chauvin2017_DiscoveryWarmDusty}, $\beta$-Pic~b \cite{lagrange2009_ProbableGiantPlanet}, AB~Aur~b \cite{currie2022_ImagesEmbeddedJovian} and PDS~70~b \cite{keppler2018_DiscoveryPlanetarymassCompanion}. These directly imaged exoplanets are represented by red diamonds in the flux ratio versus separation plot in Figure~\ref{fig:required_contrast} (right).
The best~$5\,\sigma$ detection limits of ground and space instruments are also plotted on this figure (lines). On the same plot, Jupiter-like exoplanets detected by radial velocity are shown assuming an observation in the visible light~(grey symbols). The luminosity ratio between these exoplanets and their star is~$10^{-8}$ to $10^{-9}$ in the visible. The Coronagraphic Instrument~(CGI \cite{krist2018_WFIRSTCoronagraphFlight, mennesson2020_PavingWayFuture}) aboard the future Nancy Grace Roman space telescope will aim to image these objects. Finally, the ultimate goal of direct imaging, the analysis of the atmospheres of exo-Earths  to look for biosignatures, requires an instrumental performance better than~$10^{-10}$ in luminosity ratio (dashed blue line in Figure~\ref{fig:required_contrast}, left). This is the objective of the two concept missions HabEx \cite{gaudi2020_HabitableExoplanetObservatory} and LUVOIR \cite{theluvoirteam2019_LUVOIRMissionConcept}.  A more complete description of the astrophysics results of direct imaging can be found in~\cite{chauvin2022_exoplanetimaging_CRAS,boccaletti2022_circumstellardisks_CRAS}.

This review provides a general overview of the subsystems that compose a high-contrast imaging instrument. As we cover all aspects of the process in a single paper, we sometimes refer to recent reviews that cover specific aspects in greater details. In Section~\ref{sec:coronagraphs}, we introduce the concept of stellar coronagraphy as a non-active system to suppress starlight. Section ~\ref{sec:coro_aberration} explores the sources of optical aberrations that greatly limit the performance of coronagraphs. Considering these limitations, Section \ref{sec:good_coro} defines metrics and parameters that are used to design and optimize coronagraphs. Finally in the last sections, we explain how to compensate for the effects of optical aberrations to optimize the performance of the coronagraphic system, both actively (Sections~\ref{sec:wfc},~\ref{sec:fpwfs} and~\ref{sec:experimentalvalidation}) and in post-processing once the images have been recorded (Section~\ref{sec:postproc}).

\section{Stellar coronagraphs}
\label{sec:coronagraphs}
Many methods have been suggested to obtain visible or near-infrared images of exoplanets around nearby stars, including stellar coronagraphs. Direct imaging can also be used to detect and analyze other astrophysical targets (mainly debris and protoplanetary disks). In this review, we refer to exoplanets or exoplanetary signal as any astrophysical signal that is to be detected in the environment of a star. In this section, we first explain the main challenge of high-contrast imaging (Section~\ref{subsec:why_corono}). Then, we present the concept behind stellar coronagraphs and apodizers~(Section~\ref{subsec:coro_principle}) and derive a formalism that can be used to calculate the light distribution underlying most coronagraphs~(Section~\ref{subsec:form_coro_no_aberr}). We finally discuss the manufacturing process of a coronagraph (Sections~\ref{subsec:coro_design} and \ref{subsec:example_and_fabrication_corono}).

\subsection{Why use a coronagraph?}
\label{subsec:why_corono}
\begin{figure}[!ht]
    \centering
    \includegraphics[width=\textwidth]{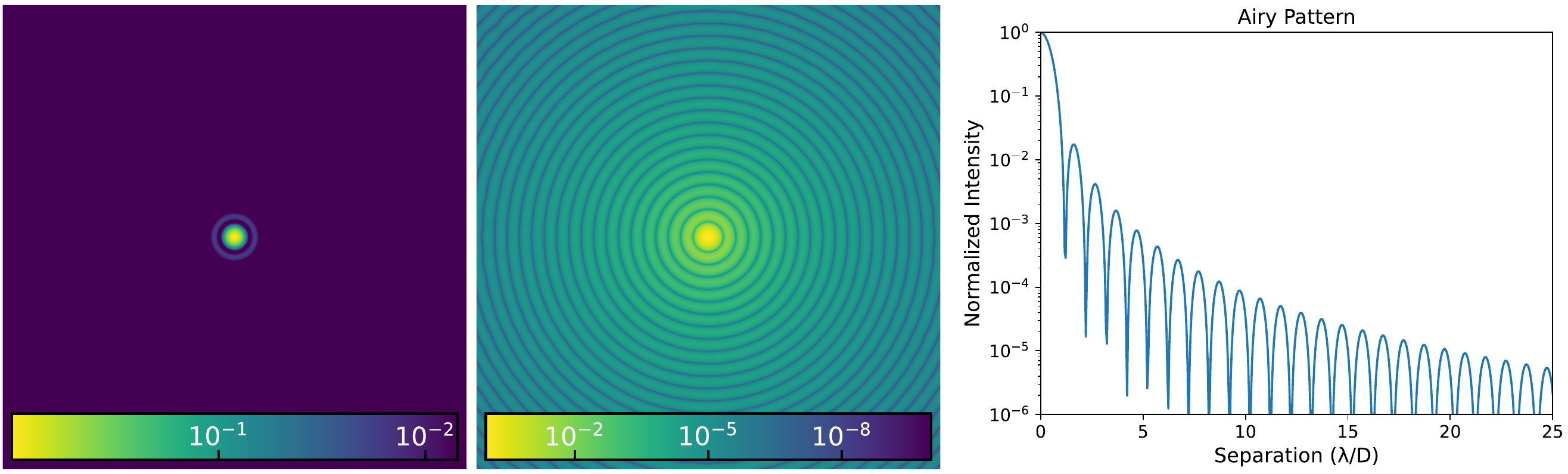}
    \caption{\it {\bf Left and center}: Normalized point spread function for a full pupil telescope of diameter~$D$  with two different color-bars. {\bf Right}: Intensity as a function of angular separation normalized to its maximum.}
    \label{fig:airy}
\end{figure}

Consider a diameter space telescope with a primary mirror diameter~$D$, associated with a detector, observing a point-like star of flux~$F_{\mathrm{S}}$, at a wavelength~$\lambda$. The image of a star through the telescope is the point spread function~(PSF) of the telescope. In a turbulence-free atmosphere, about~$80\,\%$ of the collected energy is inside a disk of~$\lambda/D$ diameter (the telescope angular resolution element, Figure~\ref{fig:airy}, left). The remaining $\sim20\,\%$ of the energy is spread over the detector, slowly decreasing with star separation (Figure~\ref{fig:airy}).
We denote by $\mathrm{PSF}\left(\vec{x}\right)$ the intensity of the star~PSF at position~$\vec{x}$ from the star center normalized to~$1$ for~$\vec{x}=\vec{0}$~(see Equation~\ref{eq:psf} for the expression of the~PSF).

Consider now an exoplanet with a flux $F_{\mathrm{P}}$, considerably fainter than that of its star $F_{\mathrm{S}}$ ($F_{\mathrm{P}}/F_{\mathrm{S}}$ ranges from $10^{-4}$ to~$10^{-10}$, Figure~\ref{fig:required_contrast}, left) and separated by a given angle in the sky. The image of the exoplanet is also a PSF (Figure~\ref{fig:airy}) but it is centered at the position~$\vec{x}$ from the star on the detector. Assuming a perfect telescope equipped with a noiseless 2D detector, the signal to noise ratio~(SNR) of the exoplanet detection $\mathrm{SNR}_{P|S}$ is the ratio between the exoplanet signal and the photon noise created by the star and exoplanet light. $\mathrm{SNR}_{P|S}$ calculated for one pixel depends on the integration time~$T_{exp}$:
\begin{equation}
    \mathrm{SNR}_{P|S}(T_{exp}) = \frac{F_{\mathrm{P}}  T_{exp}\mathrm{PSF}(0)}{\sqrt{F_{\mathrm{P}} T_{exp} \mathrm{PSF}(0) + F_{\mathrm{S}}  T_{exp}\mathrm{PSF}(\vec{x})}}
    \label{eq:init_SNR}
\end{equation}
The SNR of the same object in the absence of the star ($F_{\mathrm{S}}=0$) would be uniquely limited by the photon noise of the exoplanet light: $\mathrm{SNR}_{P} (T_{exp}) = \sqrt{F_{\mathrm{P}} \mathrm{PSF}(0) T_{exp}}$, as in \cite{pueyo2018_DirectImagingDetection}. Therefore, Equation~\ref{eq:init_SNR} can be rewritten to link the SNRs of a planet detection in the presence and absence of the star.
\begin{equation}
    \mathrm{SNR}_{P|S}(T_{exp}) = \mathrm{SNR}_{P}(T_{exp}) \left( 1+ \frac{F_{\mathrm{S}}}{ F_{\mathrm{P}} }\frac{\mathrm{PSF}(\vec{x})}{\mathrm{PSF}(0)} \right)^{-1/2}
\end{equation}

In the case of a Jupiter like planet orbiting a Sun-like star ($F_{\mathrm{P}}/F_{\mathrm{S}} = 10^{-9}$), at a projected distance of $|\vec{x}| = 5 \lambda/D$ ($\mathrm{PSF}(\vec{x})/\mathrm{PSF}(0) \simeq 10^{-3}$ for this separation, (Figure~\ref{fig:airy}, right), $\mathrm{SNR}_{P|S}(T_{exp}) = 10^{-3}\,\mathrm{SNR}_P(T_{exp})$. The presence of the star reduces the expected~SNR by a factor of~$10^3$. Hence, to reach the same~SNR, the integration time needs to be~$10^6$ times longer in the presence of the star than for the isolated exoplanet case. For a planet that could be detected at~$5\,\sigma$ in~1\,second if it were alone, $10^6\,$seconds (i.e. more than~11\,days of continuous observation) are needed for a~$5\,\sigma$ detection in the presence of the star. This represents an unreasonable amount of telescope time. Moreover, this assumes that the stellar PSF remains stable throughout the total integration time at a relative level of~$\mathrm{PSF}(\vec{x})/\mathrm{PSF}(0)$, which is far from the current stability of telescopes. Therefore, to achieve the detection of an exoplanet in a reasonable amount of telescope time, the residual starlight at the position of the exoplanet~(i.e.~$\mathrm{PSF}(\vec{x})/\mathrm{PSF}(0)$) must be minimized to values close to the flux ratio~$F_{\mathrm{P}}/F_{\mathrm{S}}$. 

Two strategies have been suggested to address this challenge: stellar coronagraphy and interferometry. Interferometers can create a destructive fringe dark enough to mask the star but narrow enough for the imaging of nearby circumstellar objects. First suggested in the 1970s \cite{bracewell1978_DetectingNonsolarPlanets}, this method was for example recently used at the Large Binocular Telescope Interferometer \cite{Hinz2017lbti} to produce the first survey of exo-zodiis \cite[debris disks equivalent to the zodiacal dust in the solar system]{ertel2020_HOSTSSurveyExozodiacal}. Another interferometric method uses the exquisite angular resolution of the Gravity/VLT interferometer to separate the fluxes from the star and planet, which are then injected into separate fibers for spectral analysis~\cite{lacour2022_interferometry_CRAS}. This method requires a good prior knowledge of the planet's position and is currently used to confirm and characterize known planets. Coronagraphy, on the other hand, aims to obtain an image of the exoplanet and is the method presented in this review.

\subsection{Coronagraph principle}
\label{subsec:coro_principle}
\begin{figure}[!ht]
    \centering
    \includegraphics[width=\textwidth]{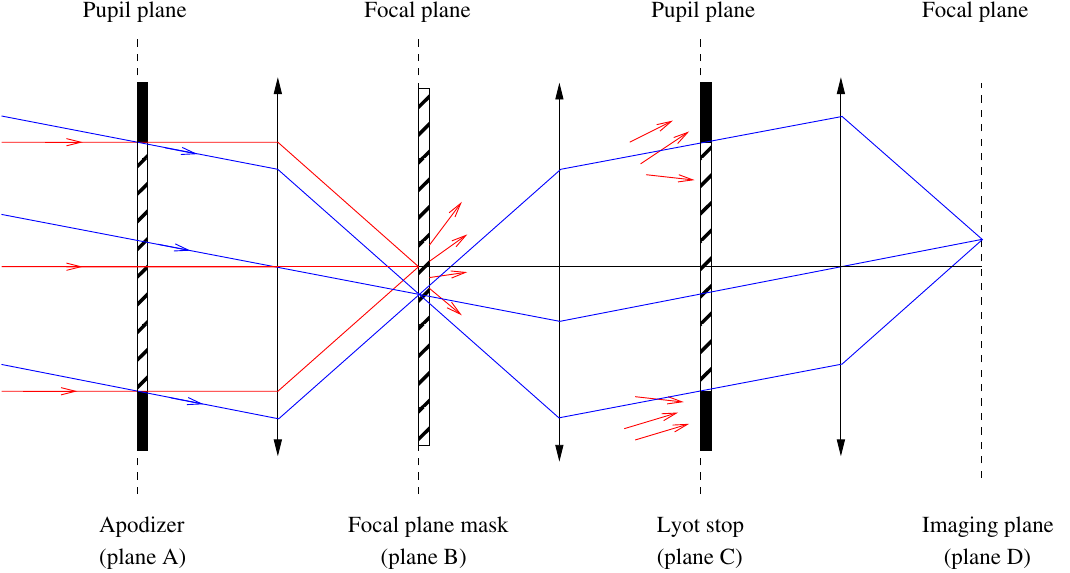}
    \caption{\it Schematic optical design of a stellar coronagraph. The starlight (red) is optically filtered whereas the exoplanet light (blue) reaches the Imaging plane. More details in the text.}
    \label{fig:principle_coro}
\end{figure}

\begin{figure}[!ht]
    \centering
    \includegraphics[width=\textwidth]{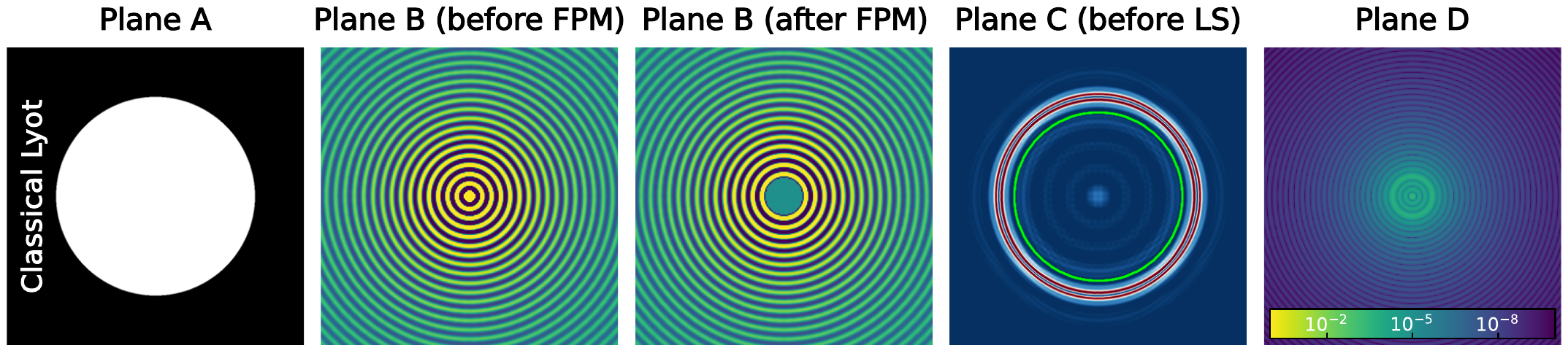}
    \caption{\it Numerical simulations of the star light distribution in planes~A to~D for a classical Lyot coronagraph. Plane A image shows the intensity in the entrance pupil of the instrument after a putative apodization~(none here). Plane~B images are the real part of the electric field before and after the focal plane mask. They share the same spatial scale and color bar. Plane~C image is the intensity distribution before the Lyot stop represented by the green circle~(all outside light is blocked). Plane~D image is the intensity distribution normalized to the maximum of the non coronagraphic~PSF. Such intensity is called normalized intensity hereafter.}
    \label{fig:coro_lyot_example}
\end{figure}

Figure~\ref{fig:principle_coro} describes the principle of a stellar coronagraph and Figure~\ref{fig:coro_lyot_example} shows the light in each plane, using the classical Lyot stellar coronagraph (as it was first suggested~\cite{kenknight1977_lyot_coro}) as an example . Other designs are described later. 

Light is collected by the telescope's primary mirror, with an entrance pupil of diameter~$D$. The optical entrance pupil of the coronagraphic system (plane~A) is optically conjugated to the pupil of the telescope. In plane~A (first image in Figure~\ref{fig:coro_lyot_example}), a pupil apodizer \cite{jacquinot1964} can be used to modify the phase and amplitude of the incoming wavefront to optimize the shape of the~diffraction pattern in the following focal plane. 
The light of the on-axis source (e.g. the star, red in the figure) is focused onto a focal plane mask in plane~B (second image of Figure~\ref{fig:coro_lyot_example}). This mask can induce spatial phase-shifts~\cite{Roddier97,Rouan00,mawet2005_AnnularGroovePhase,Soummer05, murakami08_8octant, galicher2020_FamilyPhaseMasks} and/or differential transmission~\cite{soummer2003_prolate_aplc,trauger2016_HybridLyotCoronagraph} on the electric field. In the case of the classical Lyot coronagraph, the focal plane mask is a small circular fully opaque mask. Other designs are described further in this review. The combined purpose of the pupil apodizer and the focal plane mask is to ensure that the on-axis star light is blocked and/or diffracted outside of the geometrical pupil in the following pupil plane (plane~C). In plane C, a binary diaphragm, called Lyot stop~\cite{lyot1932_EtudeCouronneSolaire}, blocks most of the remaining diffracted light (the Lyot stop is represented as a green circle in Figure~\ref{fig:coro_lyot_example}). Finally, very little stellar energy from the on-axis source reaches the final imaging plane (plane~D). The improvement can be noticed when comparing the final focal plane of the coronagraph~(Figure~\ref{fig:coro_lyot_example}, right) to the PSF in the absence of the coronagraph~(Figure~\ref{fig:airy}, center).

Conversely, the light from an off-axis source (blue lines in Figure~\ref{fig:principle_coro}) is not focused on the center of the focal plane mask in plane~B. The light therefore goes through the instrument down to the final imaging plane almost as if no mask were used in plane~A,~B and~C. As a result, the on-axis source is strongly attenuated and the off-axis source light is allowed to go through the coronagraph almost unaltered: the coronagraph acts like an optical filter that attenuates the on-axis light and transmits the off-axis light. The faint circumstellar environment (e.g. exoplanets, dust disks) of bright sources (e.g. stars) can then be imaged.

Several designs \cite{kasdin2003_ExtrasolarPlanetFinding,codona2006_app, guyon03_piaa, kenworthy2007_FirstOnSkyHighContrast,snik2012_vapp, doelman2021_VectorapodizingPhasePlate} were introduced using only one pupil apodization in plane~A and a detector, without relying on a focal plane mask to diffract light. The apodization is chosen so that the on-axis starlight is reduced within a given region of plane~B. Such instruments are usually more robust against low order aberrations than the stellar coronagraph of Figure~\ref{fig:principle_coro} but they require a high degree of accuracy in the apodization function, which can be challenging when manufacturing and aligning the masks. These designs are usually called "pupil plane coronagraphs" or "apodization only coronagraphs".
 
\subsection{Coronagraph standard formalism}
\label{subsec:form_coro_no_aberr}
In this section, a mathematical formalism is introduced to derive the distribution of light recorded by the detector in a coronagraphic instrument. Table~\ref{tab:notations} in the appendix recalls all of the notations used in this review. We use Fourier optics notions that are not detailed but we refer to specific sections in Goodman's~(2005) book \cite{goodman2005_IntroductionFourierOptics} if needed. 

Any source (star or exoplanet) is assumed to be point-like and at infinity so that the wavefront in the entrance pupil at any time~$t$ should be flat. We consider only one polarization state. We call $\omega$ the pulsation of the wave, $\vec{k}$ the wave vector, $E_{0,\lambda}$ a constant that is proportional to the square root of the incoming energy which may depend on the wavelength~$\lambda$, and $P$ the function that describes the shape of the pupil. The scalar electric field of the incoming wave in the pupil plane ~$\Psi_{\vec{k}}\left(\vec{\xi},t\right)$ at position~$\vec{\xi}$  is written as:
\begin{equation}
   \Psi_{\vec{k}}\left(\vec{\xi},t\right) = \displaystyle E_{0,\lambda}\,P\left(\vec{\xi}\right)\, e^{i\,\left(\vec{k}.\vec{\xi}-\omega\,t\right)}
   \label{eq:psi_incoming}
\end{equation}
For an on-axis source (the star), the wave vector is perpendicular to the pupil plane, hence~$\vec{k}.\vec{\xi} =0$. For an off-axis source (e.g. the planet), $\vec{k}.\vec{\xi} \ne0$.

In the following, the phasor~$e^{-i\,\omega\,t}$ is omitted, because in optical images, only intensity is recorded, which is the temporal average of the square of the real part of the electric field. Assuming a pupil apodizer called~$A_\lambda(\vec{\xi})$ is used in plane~A, the electric field~$\Psi_{A,\,\vec{k}}$ in this plane can be written as:
\begin{equation}
    \Psi_{A,\,\vec{k}}\left(\vec{\xi},t\right) = \displaystyle E_{0,\lambda}\,A_\lambda\left(\vec{\xi}\right)\,P\left(\vec{\xi}\right)\,e^{\displaystyle i\,\left(\vec{k}.\vec{\xi}\right)}
    \label{eq:psi_a_no_ab}
\end{equation}
The pupil apodizer~$A_\lambda$ modifies the phase and/or the amplitude of the incoming wave. It can have a different impact at different wavelengths~$\lambda$.

Using the Fraunhofer approximation (Section~5.3 in \cite{goodman2005_IntroductionFourierOptics}), the electric field in a focal plane is equal to the optical Fourier transform of the electric field in the previous pupil plane (Section~2.1 in \cite{goodman2005_IntroductionFourierOptics}). Hence, the electric field~$E_{B-,\,\vec{k}}(\vec{x})$ just before the focal plane mask is
\begin{equation}
    E_{B-,\,\vec{k}}\left(\vec{x}\right) = \frac{k}{2\,i\,\pi}\,\frac{e^{i\,k\,\|\vec{x}\|^2/(2\,f)}}{f^2} \iint_{\mathds{R}^2}\Psi_{A,\,\vec{k}}\left(\vec{\xi}\right)e^{\displaystyle -i\frac{k}{f}\,\vec{x}.\vec{\xi}}\mathrm{d}\vec{\xi}
\end{equation}
where $\vec{x}$ is the position in plane~B and~$f$ is the focal length of the optics that is used to go from one plane to the other. The phasor in front of the integral is usually omitted because it is close to~$1$ for usual wavelength values, pupil diameter and focal length. 
Hence, calling~$\mathrm{FT}[\Psi](\vec{u})$ the Fourier transform of the function~$\Psi$ calculated at the coordinates~$\vec{u}$, the previous equation can be written as:
\begin{equation}
   E_{B-,\,\vec{k}}\left(\vec{x}\right) \propto \FT{\Psi_{A,\,\vec{k}}}{\frac{k\,\vec{x}}{f}}
\end{equation}
This expression defines the optical Fourier transform. To simplify the equations, $\FT{\Psi}{k\,\vec{x}/f}$ is replaced by $\FT{\Psi}{\vec{x}}$ throughout the remainder of the paper to go from one pupil plane to the following focal plane. And one uses the inverse Fourier transform~$\mathrm{FT}^{-1}$ to go from one focal plane to the following pupil plane although it should be an optical Fourier transform. Hence, one changes the direction (sign) of the Cartesian coordinates after two optics (from one pupil plane to the next one).

The intensity distribution in plane~B before the focal plane mask~$M$ is the temporal average of the square of the real part of the electric field. Accounting for the phasor~$e^{-i\,\omega\,t}$ and after calculation, the intensity can be written as:
\begin{equation}
   I_{B-,\,\vec{k}}\left(\vec{x}\right)\propto\left|E_{B-,\,\vec{k}}\left(\vec{x}\right)\right|^2
   \label{eq:psf_non_norm}
\end{equation}
It is also the~PSF of the instrument, represented in Figure~\ref{fig:airy} once it is normalized to its maximum
\begin{equation}
    \mathrm{PSF}\left(\vec{x}\right)=\frac{I_{B-,\,\vec{k}}\left(\vec{x}\right)}{\mathrm{max}\left(I_{B-,\,\vec{k}}\right)}
    \label{eq:psf}
\end{equation}
The electric field in plane~B then encounters the focal plane mask~$M_\lambda$:
\begin{equation}
    E_{B,\,\vec{k}}\left(\vec{x}\right) \propto \FT{\Psi_{A,\,\vec{k}}}{\vec{x}}\,M_\lambda(\vec{x})
\end{equation}
This mask can be a phase mask ($M_\lambda$ is a phasor that modifies phase of the incoming wavefront only), an amplitude mask ($M_\lambda$ is a real function that modifies amplitude only) or a combination of both. Examples of such masks are presented in section~\ref{subsec:example_and_fabrication_corono}. The field~$\Psi_{C,\,\vec{k}}$ after the Lyot stop~$L_\lambda(\xi)$ can be calculated using the inverse Fourier transform~$\mathrm{FT}^{-1}$ (we assume no magnification between planes~A and~C)
\begin{equation}
    \Psi_{C,\,\vec{k}}\left(\vec{\xi}\right) \propto L_\lambda\left(\vec{\xi}\right)\,\FTinv{E_{B,\,\vec{k}}}{\vec{\xi}}
\end{equation}
This equation can also be written using a convolution product denoted with the symbol~$\star$:
\begin{equation}
    \Psi_{C,\,\vec{k}}\left(\vec{\xi}\right) \propto L_\lambda\left(\vec{\xi}\right)\,\left(\Psi_{A,\,\vec{k}}\star\mathrm{FT}^{-1}\left[M_\lambda\right]\right)\left(\vec{\xi}\right)
    \label{eq:psi_c_no_ab}
\end{equation}
As for the apodizer~$A_\lambda$ and the focal plane mask~$M_\lambda$, the Lyot stop~$L_\lambda$ can modify the phase~(rare but possible) and/or the amplitude of the incoming wave and it can depend on the wavelength~$\lambda$. Still assuming Fraunhofer propagation, and assuming no magnification between planes~B and~D, the electric field~$E_{D,\,\vec{k}}$ in the final imaging plane~D can be written as:
\begin{equation}
   E_{D,\,\vec{k}}\left(\vec{x}\right) \propto \FT{\Psi_{C,\,\vec{k}}}{\vec{x}}
\end{equation}
Using~Eqs.~\ref{eq:psi_a_no_ab} and~\ref{eq:psi_c_no_ab},~$E_{D,\,\vec{k}}$ can be written as a function of the incoming beam properties ($E_{0,\lambda}$ and $\vec{k}$):
\begin{equation}
   E_{D,\,\vec{k}}\left(\vec{x}\right) \propto \mathrm{FT}\left[L_\lambda\right]\star\left(\mathrm{FT}\left[E_{0,\lambda}\,A_\lambda\left(\vec{\xi}\right)\,P\left(\vec{\xi}\right) e^{\displaystyle i\,\vec{k}.\vec{\xi}}\right]\,M_\lambda\right)\left(\vec{x}\right)
   \label{eq:ED_no_ab}
\end{equation}
Following Give'on et al. (2007) \cite{giveon2007_ClosedLoopDM}, we introduce a linear operator $\mathcal{C}$ that links the electric field~$E_{D,\,\vec{k}}$ in the imaging plane~D and the electric field~$\Psi_{\vec{k}}$ in the entrance pupil plane~A: 
\begin{equation}
   E_{D,\,\vec{k}} \propto \mathcal{C} \left[\Psi_{\vec{k}}\right] = \mathrm{FT}\left[L_\lambda\right]\star\left(\mathrm{FT}\left[A_\lambda \, \Psi_{\vec{k}}\right]\,M_\lambda\right)
   \label{eq:C_operator2}
\end{equation}
Finally, the recorded intensity in plane~D can be written as:
\begin{equation}
    \label{eq:intensity_corono}
   I_{D,\,\vec{k}}\left(\vec{x}\right) \propto \left|E_{D,\,\vec{k}}(\vec{x})\right|^2
\end{equation}
The detector used for detecting the exoplanet's signal is in this plane, often called the science plane. The previous formula gives the recorded intensity for a unique monochromatic source at~$\lambda=2\,\pi/k$. If several sources are in the field of view, the recorded intensity is the sum of the intensities calculated for each individual source (the lights coming for the different sources are incoherent). For example, consider a star on the optical axis~($\vec{k}.\vec{\xi}=0$ in Equation~\ref{eq:psi_incoming}) with~$I_{D,\,\mathrm{S}}$ its intensity calculated from~Eq.~\ref{eq:intensity_corono} and, an off-axis source like an exoplanet~($\vec{k}.\vec{\xi}\ne0$) with~$I_{D,\,\mathrm{P}}$ its intensity. In such a case the recorded intensity~$I_D$ is given by:
\begin{equation}
I_D(\vec{x},\lambda,p) = I_{D,\,\mathrm{S}}(\vec{x},\lambda,p)+ I_{D,\,\mathrm{P}}(\vec{x},\lambda,p)
\label{eq:ID_star_planet}
\end{equation}
where we add the putative dependence of the intensity with the polarization state~$p$. The reader might notice that the distribution of light in the coronagraphic image (plane~D) is not the convolution of the astronomical scene (here, two point-like sources) by the {\it PSF} of the instrument because the~{\it PSF} strongly varies in the field of view (close to~$0$ on the optical axis and almost unaffected far from this axis). Because this property of convolution of the scene by a uniform PSF is often presented as an important property in astronomical imaging, it is best to avoid using the term {\it PSF} for coronagraphic images. 

For a given spectral filter centered on~$\lambda$ and of bandwidth~$\Delta\lambda$, the recorded intensity~$I_{D,\Delta\lambda}$ is the integration of the monochromatic intensity over the filter. Furthermore, if two orthogonal polarization states~$p$ are considered, the total intensity is the sum of the intensities calculated for each state of polarization. Finally, the recorded intensity can be written as:
\begin{equation}
    \label{eq:intensity_corono_bandwidth}
   I_{D,\Delta\lambda}\left(\vec{x}\right) \propto\sum_p \int_{\lambda-\Delta\lambda/2}^{\lambda+\Delta\lambda/2} I_D(\vec{x},\lambda,p)\,\mathrm{d}\lambda
\end{equation}

\noindent Hereafter, we consider a monochromatic case and a single polarization state (up to Eq.~\ref{eq:ID_star_planet}).

\subsection{How to design a coronagraph?}
\label{subsec:coro_design}
The goal of the stellar coronagraph is to:
\begin{itemize}
    \item minimize~$I_{D,\,\vec{k}}=I_{D,\,\mathrm{S}}$ ($S$ for star) for the on-axis source ($\vec{k}.\vec{\xi}=0$ in Equation~\ref{eq:psi_incoming}) ;
    \item maximize~$I_{D,\,\vec{k}}=I_{D,\,\mathrm{P}}$ ($P$ for planet) for off-axis sources ($\vec{k}.\vec{\xi}\ne0$).
\end{itemize}
The free parameters are the three masks~$A_\lambda$ (apodizer), $M_\lambda$ (focal plane mask) and~$L_\lambda$ (Lyot stop) which can modify the wavefront phase, amplitude or both waves. In an ideal case, the stellar energy is totally stopped and the energy of the exoplanet is totally transmitted. Using Equation~\ref{eq:ED_no_ab}, the two conditions can be written as:
\begin{subequations}
\begin{empheq}[left=\empheqlbrace]{align}
   &\mathrm{FT}\left[L_\lambda\right]\star\left(\mathrm{FT}\left[A_\lambda\,P\right]\,M_\lambda\right)\left(\vec{x}\right)=0\qquad & \mathrm{for}\ \mathrm{any}\ \vec{x}\label{eq:coro_on_axis}\\
   &I_{D,\,\vec{k}}\left(\vec{x}\right) = I_{B-,\,\vec{k}}\left(\vec{x}\right) \propto \left|\FT{E_{0,\lambda}\,A_\lambda\left(\vec{\xi}\right)\,P\left(\vec{\xi}\right)\,e^{\displaystyle i\,\vec{k}.\vec{\xi}}}
   {\vec{x}}\right|^2\qquad & \mathrm{for}\ \vec{k}.\vec{\xi}\ne0 \label{eq:coro_off_axis}
\end{empheq}
\end{subequations}
Such an ideal coronagraph cannot be built because if Equation~\ref{eq:coro_on_axis} is verified, part of the off-axis electric field is modified by the coronagraph and Equation~\ref{eq:coro_off_axis} cannot be verified~\cite{guyon2006_limit_coro}. 

\begin{figure}[!ht]
    \centering
    \includegraphics[width=0.8\textwidth]{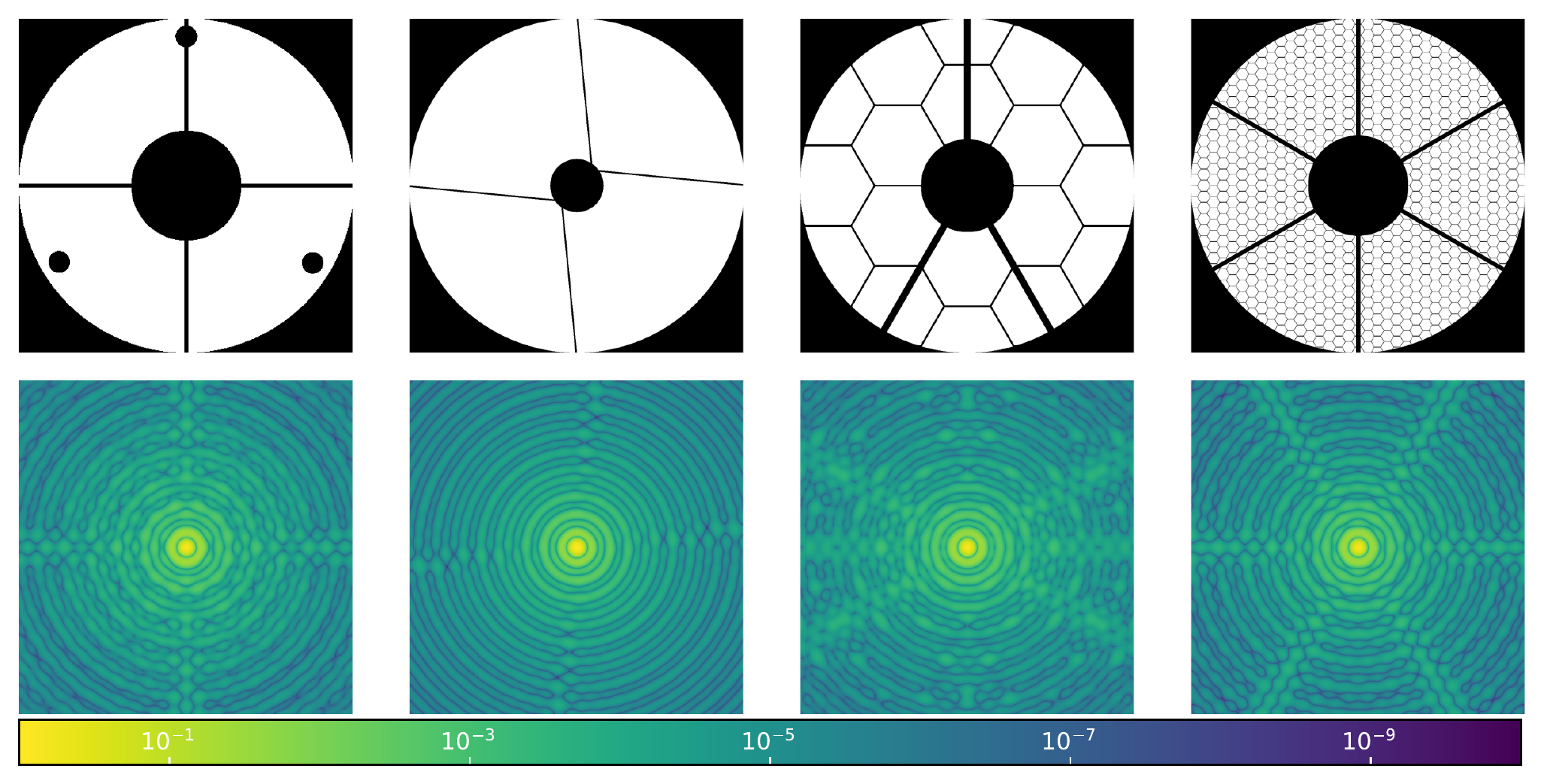}
    \caption{\it {\bf Left to Right}: Apertures (top) and associated normalized intensity of the PSF (bottom) for the HST, VLT, JWST and ELT (in plane~B). For the ELT and JWST the inner and outer edge of the pupil were made circular which is customary when designing coronagraphs. Apertures were re-created using \cite{por2018hcipy}.}
    \label{fig:complex_apertures}
\end{figure}

Numerous stellar coronagraphic designs have been suggested to solve Equation~\ref{eq:coro_on_axis}. Some have been proven to theoretically cancel the star light in the whole focal plane~D \cite{Rouan00, mawet2005_AnnularGroovePhase, soummer2003_prolate_aplc, Hou2014Wide-band, galicher2020_FamilyPhaseMasks} whereas others are optimized to minimize the star light in part or the totality of the focal plane~D \cite{Roddier97,kuchner2002_CoronagraphBandlimitedMask,guyon03_piaa} and maximize the transmission of the planet image.

The first stellar coronagraphs were optimized for full circular aperture as for an off-axis telescope~\cite{kenknight1977_lyot_coro}, but these designs are severely limited for more complex telescope apertures \cite{sivaramakrishnan2005_LyotCoronagraphyGiant}. For example, on-axis telescopes have central obscurations, spiders and putative segmentation that scatter the star light in the focal plane~B of the coronagraph, as shown in Figure~\ref{fig:complex_apertures}. These diffraction patterns usually strongly degrade the performance of coronagraphs designed for clear circular apertures.
For two decades, coronagraph solutions have been suggested for more complex telescope apertures, often using optimized apodizations~$A_\lambda$. Designs were first proposed to achieve high starlight suppression with a central secondary obscuration ~\cite{Soummer05, mawet2013_RingapodizedVortexCoronagraphs} and for any given apertures, including spiders and/or segmentation~\cite{enya2010,carlotti2013_ApodizedPhaseMask, guyon2014_HighPerformanceLyot, haze2015,ruane2015_LyotplanePhaseMasks, trauger2016_HybridLyotCoronagraph, zimmerman2016_ShapedPupilLyot, ndiaye2018_complex_aperture_aplc,pueyo2013_HighcontrastImagingArbitrary,por2020_paplc}. A major remaining hurdle is the loss of signal in the core of the exoplanet image as the secondary obscuration increases~\cite{fogarty2017_PolynomialApodizersCentrally}.

\subsection{Examples and fabrication of apodizers, focal plane masks and Lyot stops}
\label{subsec:example_and_fabrication_corono}
\begin{figure}[!ht]
    \centering
    \includegraphics[width=\textwidth]{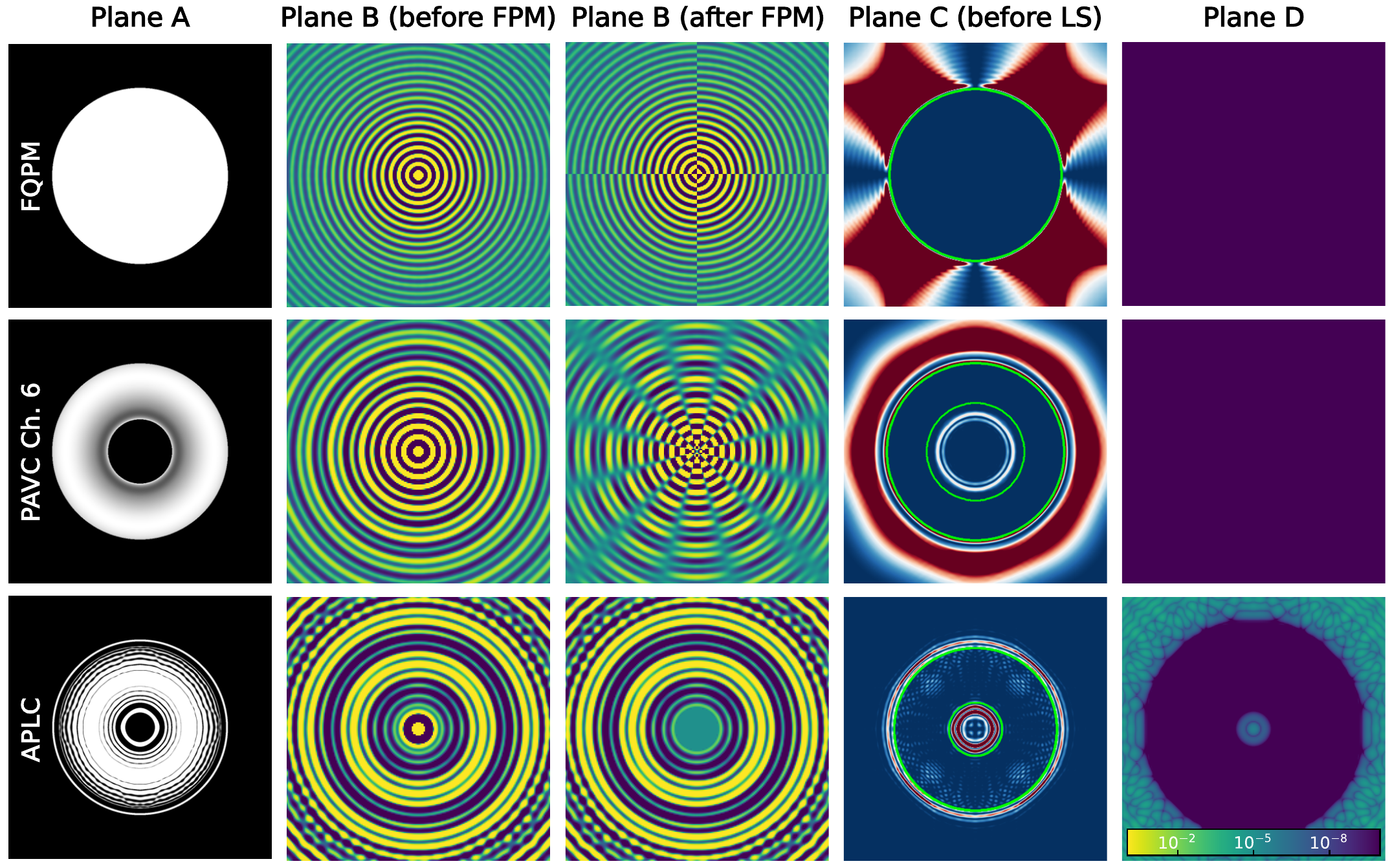}
    \caption{\it Same as Figure~\ref{fig:coro_lyot_example} for three other coronagraphs. \textbf{From top to bottom:} Numerical simulations of the light distribution in plane~A to plane~D for a Classical Lyot coronagraph; a four-quadrant phase mask~\cite[FQPM]{Rouan00}; a Polynomial apodized Vortex coronagraph \cite{fogarty2017_PolynomialApodizersCentrally} of charge 6 designed for a central obscuration of~$36\,\%$ of the pupil ; an APLC \cite{ndiaye2018_complex_aperture_aplc} designed for a central obscuration of~$17\,\%$ of the pupil.}
    \label{fig:coro_example}
\end{figure}

Many stellar coronagraph designs have been suggested and we choose not to review all of them. Several reviews ~\cite{guyon2006_limit_coro,mawet2012_ReviewSmallangleCoronagraphic,ruane2018_ReviewHighcontrastImaging} detail the different families of coronagraphs and their respective advantages. In Figure~\ref{fig:coro_example}, we show only a few examples of coronagraphs so that the reader can {\it see} the effect of each of the three masks~$A_\lambda$ (apodizer in plane A), $M_\lambda$ (focal plane mask, FPM in plane B) and~$L_\lambda$ (Lyot stop, LS in plane C). 

First, we consider a four-quadrant phase mask coronagraph~\cite[FQPM]{Rouan00}. In plane A, there is no apodization~($A_\lambda=1$). The focal plane mask~$M_\lambda$ induces a $\pi-$phase shift on two quadrants in a diagonal with respect to the two others. The effect is visible on the electric field in the focal plane~B. Contrarily to the classical Lyot coronagraph (Figure~\ref{fig:coro_lyot_example}), the stellar light is not blocked but only phase-shifted in the focal plane. The phase-shift is such that all the starlight is diffracted outside the geometrical pupil and stopped by the Lyot stop in plane~C. No starlight reaches the detector (plane~D) (Equation~\ref{eq:coro_on_axis} is respected in theory). For example, FQPM coronagraphs are embedded in the MIRI instrument \cite{boccaletti2015_MidInfraredInstrumentJames} aboard JWST.

The second example is a polynomial apodized vortex coronagraph \cite[PAVC] {fogarty2017_PolynomialApodizersCentrally} of charge 6. This coronagraph was designed to cancel the diffraction created by a central obscuration. This central obscuration superimposed on the amplitude apodization function~$A_\lambda$ is shown in plane A. The focal plane mask introduces an azimuthal phase ramp going from~0 to~$12\,\pi$ radians. In this case, the light is diffracted either in the outer region of the plane C or in the center of the plane C. It is stopped by a centrally obscured Lyot stop for which the boundaries are represented by two green circles. This coronagraph obeys Equation~\ref{eq:coro_on_axis} in theory and no star light reaches the detector (plane~D). 

Finally, the last example is an apodized pupil Lyot coronagraph (APLC) designed to cancel the diffraction created by a central obscuration~\cite{ndiaye2018_complex_aperture_aplc}. The amplitude apodization function~$A_\lambda$ concentrates the energy behind the focal plane mask~$M_\lambda$ in plane~B. The focal plane mask is an opaque disk. The light is mainly diffracted either in the outer region of the plane C or in the center of the plane C, where it is mostly stopped by a centrally obscured Lyot stop represented by two green circles. There is residual starlight on the detector (plane~D): this coronagraph does not obey Equation~\ref{eq:coro_on_axis}. However, the apodization for this coronagraph is optimized to minimize the starlight below a certain level in a given region of the focal plane~D (dark region in the image on the right). These types of designs are the main coronagraphs of both VLT/SPHERE and Gemini/GPI instruments \cite{boccaletti2008_PrototypingCoronagraphsExoplanet, sivaramakrishnan2010_GeminiPlanetImager}.

Once the functions~$A_\lambda$, $M_\lambda$ and~$L_\lambda$ have been defined (analytically or empirically), the fabrication of the optical components can start. There are many possibilities to create phase masks, that only change the phase of the electric field, amplitude masks, that only modify the modulus of the field, or {\it complex} masks that modify both phase and amplitude.

For amplitude masks, historically binary masks have been explored \cite{debes2004_binary_mask_lab,chakraborty2005_binary_mask_lab} but other solutions exist, such as deposits of microdots~\cite{martinez2009_microdot1}, or even micro-mirrors~\cite{carlotti2018_micromirror,kagitani2020_micromirror}. For phase masks, manufacturing solutions include steps of materials~\cite{bonafous2016_step_material_FQPM,ruane2019_scalar_vortex,galicher2020_FamilyPhaseMasks}, liquid crystal polymers~\cite{mawet2011_subwavelength_grating_liq_crystal,murakami2013_photonic_crystal,doelman2020_liquid_crystal}, sub-wavelength gratings~\cite{mawet2005_AnnularGroovePhase,niv2007_subwavelength_grating,mawet2011_subwavelength_grating_liq_crystal,delacroix2012_subwavelenght_grating} or, birefringent materials~\cite{boccaletti2008_HW_FQPM}. Each technical solution has its advantages and disadvantages: more or less easy to fabricate, more or less chromatic or polarized, etc. Finally, apodization can also be designed with several cascading optics in cascade of optimized shapes~\cite{guyon03_piaa,belikov2006_apod_pupil_several_optics}. 

\section{Optical Aberrations}
\label{sec:coro_aberration}

\subsection{Coronagraph formalism with aberrations}
\label{subsec:corono_with_aberr}

In the formalism presented in the Section~\ref{subsec:form_coro_no_aberr}, it was specifically assumed a point-source object located at infinity resulting in a perfectly flat wavefront in the entrance pupil of the coronagraph~(plane~A), scattering into an ideal~PSF in the following focal plane (shown in Figure~\ref{fig:airy}). This is usually referred to as the diffraction limited regime. We now consider the effects of aberrations, resulting in a non-perfectly flat wavefront entering the coronagraphic system.

The first kind of aberrations considered in this paper are phase aberrations, i.e. delays or advances of part of the wavefront with respect to a flat wavefront. Phase aberrations can be introduced by the Earth's atmosphere for telescopes on the ground, and/or by manufacturing imperfections in the telescope's optics (reflective or refractive). In the context of coronagraphy, amplitude aberrations must be considered too. They are local transmission differences over the beam. They can be caused by small holes, dust or coating defects on the optics. It should be noted that a phase aberration introduced by an optic that is not conjugated with the pupil plane can result in a mix of phase and amplitude aberrations in the pupil plane, due to the Fresnel propagation (the process is detailed in Section~\ref{subsec:corr_2DM}). Usually, in coronagraphy, all phase and amplitude aberrations on the wavefront introduced in any plane before the focal plane~B are described by a single phase~$\phi\iup(\vec{\xi},t)$ aberration term and a single amplitude~$a\iup(\vec{\xi},t)$ aberration term in the entrance pupil plane~A. Equation~\ref{eq:psi_incoming} can then be written as:
\begin{equation}
    \Psi_{\vec{k}}\left(\vec{\xi},t\right) = \displaystyle E_{0,\lambda}\,P\left(\vec{\xi}\right)\,\,e^{\displaystyle a\iup(\vec{\xi},t)+i\,\phi\iup(\vec{\xi},t)}
    \label{eq:psi_incoming_aber}
\end{equation}
Similarly, all phase and amplitude aberrations introduced in any plane after the focal plane~B are described by a single phase~$\phi\ido(\vec{\xi})$ aberration term and a single amplitude~$a\ido(\vec{\xi})$ aberration term in the Lyot stop pupil plane~C. Equation~\ref{eq:psi_c_no_ab} becomes
\begin{equation}
    \Psi_{C,\,\vec{k}}\left(\vec{\xi},t\right) \propto L\left(\vec{\xi}\right)\,e^{\displaystyle a\ido(\vec{\xi},t)+i\,\phi\ido(\vec{\xi},t)}\,\left(\Psi_{A,\,\vec{k}}\star\mathrm{FT}^{-1}\left[M\right]\right)\left(\vec{\xi}\right)
    \label{eq:psi_c}
\end{equation}
Consequently, assuming phase and amplitude aberrations, and an on-axis source, Equation~\ref{eq:ED_no_ab} can be written as:
\begin{equation}
   E_{D,\,\mathrm{S}}\left(\vec{x},t\right) \propto \mathrm{FT}\left[L\left(\vec{\xi}\right)\,e^{\displaystyle a\ido(\vec{\xi},t)+i\,\phi\ido(\vec{\xi},t)}\right]\star\left(\mathrm{FT}\left[E_{0,\lambda}\,A(\vec{\xi})\,e^{\displaystyle a\iup(\vec{\xi},t)+i\,\phi\iup(\vec{\xi},t)}\right]\,M\right)\left(\vec{x}\right)
   \label{eq:ED}
\end{equation}
This equation shows that, for a given triplet of masks~$A_\lambda$, $M_\lambda$ and~$L_\lambda$, the performance depends on the aberrations. Because they are introduced after the coronagraphic focal plane mask, downstream aberrations have a limited impact on the coronagraph's capacity to diffract and block the starlight. We neglect them in this review and more information can be found in the literature~\cite{paul2013_CoronagraphicPhaseDiversity, herscovici-schiller2018_ExperimentalValidationJoint}.

Using the linear operator introduced in Equation~\ref{eq:C_operator2}, the electric field in the imaging plane of the coronagraph is given by: 
\begin{equation}
  E_{D,\,\mathrm{S}} = \mathcal{C}\left[ P\, e^{\displaystyle a\iup + i\,\phi\iup} \right] = \mathcal{C}\left[ P \right] + \mathcal{C}\left[ P\, \left(e^{\displaystyle a\iup + i\,\phi\iup} -1 \right) \right]
\label{eq:ephiminus1}
\end{equation}
The first term, $\mathcal{C}\left[ P \right]$, is the response of the coronagraph to the telescope aperture for the case without aberration. This is the "known static part" and coronagraph designs can be optimized in advance to minimize or cancel this term (Equation \ref{eq:coro_on_axis}), even for apertures with central obscurations and discontinuities (Section~\ref{subsec:coro_design}). The second term due to aberrations is the "unknown part" of the stellar intensity. It appears in the final focal plane~D as speckles, shown in Figure~\ref{fig:PSF_coroIm_aberration} (right). These stellar speckles mimic point-like source images (e.g. exoplanet images) and severely limit the performance of coronagraphs. Their position and intensity change with changes in the aberrations. This "unknown part" can be minimized using active optics during the observation with wavefront control techniques~(Sections~\ref{sec:wfc},~\ref{sec:fpwfs} and~\ref{sec:experimentalvalidation}). It can also be calibrated after the observation using adapted observational strategies and post-processing techniques (Section~\ref{sec:postproc}).

The distinction between "known part" and "unknown part" is now blurring, especially in the design of coronagraphs for non-clear apertures (central obscuration, spiders, segmentation). These instruments sometimes purposely include coronagraphs that are not fully canceling the "known term" on purpose ($|\mathcal{C}\left[ P \right]|^2 \ne 0$), and rely on active systems to simultaneously minimize the diffraction created by the pupil discontinuities and the speckles induced by unknown aberrations\cite{pueyo2013_HighcontrastImagingArbitrary, mazoyer2018_ActiveCorrectionApertureI,trauger2016_HybridLyotCoronagraph}. Conversely, coronagraphs are now designed not only to cancel the known term but also specifically to minimize low order aberrations~\cite{fogarty2020_HighThroughputLoworder}. In the remainder of the review, we assume that Equation~\ref{eq:coro_on_axis} is verified: the coronagraph design cancels the "known part"~($\mathcal{C}\left[ P \right] = 0$) and we focus on the second term of Equation~\ref{eq:ephiminus1}. 

\subsection{Dynamic, quasi-static or static aberrations and speckles}
Speckle evolution has been studied for ground-based coronagraphic instruments \cite{hinkley2007_TemporalEvolutionCoronagraphic,martinez2013_SpeckleTemporalStability, males2021_Mysterious_Lives_of_Speckles, vigan2022_CalibrationQuasistaticAberrationsa} and HST \cite{lallo2006_TemporalOpticalBehavior} and estimated for the Nancy Grace Roman Space Telescope using thermal and structural modeling \cite{krist2016_NumericalModelingProposed}.

The speckle intensity and position change in the coronagraphic image as the phase and amplitude aberrations vary. If the integration time is longer than the speckle lifetime, the resulting image is the average over several speckle patterns. Understanding the temporal evolution of the optical aberrations relative to the integration time is thus critical when designing an instrument to optimize the active and the \textit{a posteriori} calibrations. 
\begin{figure}[!ht]
    \centering
    \includegraphics[width=.8\textwidth]{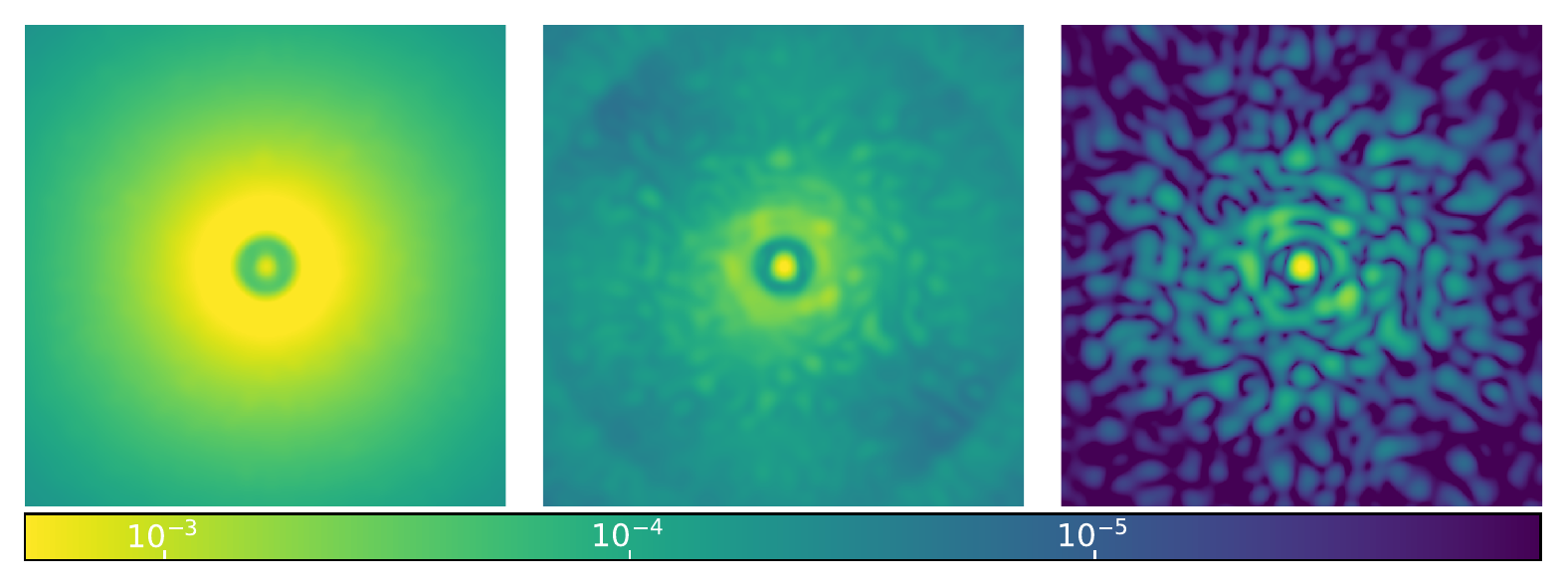}
    \caption{\it Numerical simulations of the coronagraph focal plane~D in the presence of aberrations. All simulations assume an APLC in a SPHERE-like system for a 30 s observation. The color bar gives the normalized intensity value. \textbf{Left:} Ground-based instrument with no AO system. The rapidly varying speckles create a smooth halo and the coronagraph is rendered useless. \textbf{Center:} Ground-based instrument with an AO system. Most of the speckles are corrected and only a smooth halo remains due to AO residuals. Quasi-static speckles are present above the halo. \textbf{Right:} Ground-based instrument with an ideal~AO system removing all dynamic speckles, or a space-based instrument. Only quasi-static speckles are visible.}
    \label{fig:PSF_coroIm_aberration}
\end{figure}

For example, the  main source of aberrations for ground-based coronagraphic instruments is the Earth's atmosphere. It creates an important level of aberrations, with optical path differences of the order of the wavelength for the visible and near-IR ($\sim1\,\mu$m). These aberrations also vary on the millisecond timescale, much faster than the integration time of the science detector located in the focal plane of the coronagraph. This integration time is driven by the magnitude of the observed astrophysical object and is usually in the range of a few seconds to a few tens of second. Speckles faster than the integration time are called dynamic speckles. The resulting coronagraphic image is the average of thousands of speckle patterns. The star light appears in the science image not as individual speckles but as  a smooth halo decreasing from the center of the image and much brighter that the exoplanet image. The images in Figure~\ref{fig:PSF_coroIm_aberration} show numerical simulations of the effect of aberrations introduced by the Earth's atmosphere on the coronagraph science plane. The left image shows the smooth halo of an uncorrected ground-based coronagraph. In Section~\ref{subsec:adaptive_opt}, we discuss adaptive optics (AO) techniques that compensate most of the fast atmospheric aberrations. In the center of figure~\ref{fig:PSF_coroIm_aberration}, most of the smooth halo have been removed thanks to the AO correction. However, classical~AO systems minimize the phase aberration measured by a wavefront sensor~(WFS), located in an optical channel physically separated from the coronagraph channel. Therefore, the aberrations corrected by the AO are not fully identical to the aberrations seen by the coronagraph and there are non-common path aberrations between the two channels. These aberrations can evolve slower than the exposure time and induce speckles above the halo as in the center of Figure~\ref{fig:PSF_coroIm_aberration}. From space, only slowly evolving aberrations exist and the current coronagraphic image is dominated by speckles~(right).

Speckles that remain unchanged during the whole observation sequence are static. They can be calibrated by post-processing techniques (Section~\ref{sec:postproc}) or minimized on-sky (Sections~\ref{sec:wfc} to~\ref{sec:experimentalvalidation}). Speckles stable for only a few exposures of the observation sequence are called quasi-static. These speckles are particularly problematic for exoplanet imaging. First of all, their shape is similar to off-axis point-like source images (e.g. exoplanet images) inducing false detection. Furthermore, because they evolve during the observation sequence, they are much more resistant to post-processing techniques. One solution is the use of an active correction during the observation to minimize their intensity but here again, the measurement and correction of the speckles must be faster than the speckle lifetime. A formalism for quasi-static speckles is presented in Section~\ref{subsec:quasi_static} and the techniques for their sensing and correction are introduced in Sections~\ref{sec:wfc}~to~\ref{sec:experimentalvalidation}. 

These categories of static, quasi-static and dynamic speckles are often used to understand or predict the performance of instruments. However, it must be kept in mind that there is a continuum of lifetimes for speckles for a given instrument which means that these categories depend on the exposure time. 

The expression {\it speckle noise} is often incorrectly used to describe any speckle. However, long lifetime speckles can be corrected (sections~\ref{sec:wfc} to~\ref{sec:experimentalvalidation}) or calibrated \textit{a posteriori}~(Section~\ref{sec:postproc}). They should be referred as 'bias', not 'noise'. We encourage the reader to use 'speckles' or 'speckle pattern' for such speckles. Conversely, {\it Speckle noise} can refer to shorter lifetime speckles that change in intensity and position from one frame to the other. In this case, one can decrease the exposure time to render these speckles quasi-static, or increase the exposure time to average the speckles (creating a stellar halo that adds photon noise and limits the signal-to-noise of exoplanet detection, see Equation~\ref{eq:snr_speckle_and_photon_noise}).

\subsection{Dynamic aberrations and adaptive optics}
\label{subsec:adaptive_opt}

Although the first coronagraphs were installed on ground-based telescopes in the 1980's (detection of the $\beta$-Pictoris debris disk in~1984 at Las Campanas Observatory \cite{smith1984_CircumstellarDiskBeta}), the first exoplanet images were obtained in the late~2000's~\cite{marois2008_DirectImagingMultiple,lagrange2009_ProbableGiantPlanet}. These detections were only made possible once adaptive optics systems~(AO) were installed to compensate for atmosphere-induced phase aberrations~(VLT/NACO, Keck/NIRC2, NIRI/Gemini North, etc). The~AO system requires a specific optical channel in the instrument for measuring the aberrations to be corrected using a~WFS usually working at a different wavelength than the one of the light sent into the coronagraph. The field of techniques developed for wavefront sensing and correction of atmospheric aberrations is wide. Its applications in astronomy are much wider than just coronagraphy (although because of the extremely high performance required, stellar coronagraphy is a clear driving force behind the development of some of the high-performance single conjugated~AO systems). A description of~AO systems and of the~AO performance is out of the scope of this review. We refer the reader to the book written by~Roddier~\cite{roddier1999_AdaptiveOpticsAstronomy} or, more recently and more concisely the reviews Guyon (2018) \cite{guyon2018_ExtremeAdaptiveOptics} (in English), and Rousset \& Fusco (2022) \cite{rousset2022_AO_CRAS} (in French).

In current instruments and under good atmospheric conditions, the coronagraphic image is the sum of a smooth halo due to dynamic aberrations and quasi-static speckles induced by quasi-static aberrations (Figure~\ref{fig:PSF_coroIm_aberration}, center).
Several authors have analyzed the effect of turbulence on coronagraphic images with and without~AO~\cite{perrin2003_StructureHighStrehl,sauvage2010_AnalyticalExpressionLongexposure}. More recently, formalisms have been published to predict the light distribution in images the in coronagraphic focal plane of extreme-AO for long exposures \cite{herscovici-schiller2017_AnalyticExpressionCoronagraphic,singh2019_long_exposure} or to reconstruct coronagraphic images from~AO telemetry~\cite{guyon2020_psf_reconstruction}. Finally, the coronagraphic AO~residuals (temporal average of the square of second terms of Equation~\ref{eq:ephiminus1}) is impacted by many factors \cite[atmospheric dispersion, diffraction effects, low wind effect, and so on]{cantalloube2019_PeeringSPHEREImages}. 
The~AO system correction sets a theoretical limit on the best normalized intensity that can be reached from the ground at~$\sim10^{-7}$ within the central arc-second for~8m class telescopes, which can be improved to~$\sim10^{-8}$ using post-processing techniques~\cite[Section~\ref{sec:postproc}]{guyon2005_LimitsAdaptiveOptics, fusco2006_DesignExtremeAO, guyon2017_AdaptiveOpticsPredictive},

Among the dynamic aberrations, we usually single out low order aberrations that spatially vary in a pupil plane with low spatial frequencies. They correspond to starlight leakage close to the optical axis in the coronagraphic image. These aberrations are both the ones with the highest energy and the most critical to probe the region at a few angular resolution elements from the star where most exoplanets still hide. These low order aberrations are also present in space-based instruments. For example, recent studies have evaluated them for the Roman Space Telescope~\cite{shi2017_DynamicTestbedDemonstration}. To stabilize or correct for these aberrations, a whole class of low-order wavefront sensing (LOWFS) techniques have been developed. Among these we can cite techniques that study the distribution of light in the coronagraphic images~\cite{mas2012_TiptiltEstimationCorrection,huby2015_PostcoronagraphicTiptiltSensing}, the~LOWFS techniques that uses light rejected outside of the Lyot stop~\cite{singh2014_LyotbasedLowOrder} and, the Zernike~WFS \cite{zernike1934_DiffractionTheoryKnifeedge, wallace2012_PhaseshiftingZernikeWavefront, ndiaye2013_CalibrationQuasistaticAberrations, vigan2019_CalibrationQuasistaticAberrations,shi2017_DynamicTestbedDemonstration}. The two latter introduce phase shifting optics within the beam or record out-of-focus images to break the phase degeneracy (Section \ref{subsec:phase_degeneracy}).

\subsection{Quasi-static aberrations}
\label{subsec:quasi_static}
In the remainder of the review, we will only consider quasi-static aberrations: they are static during one exposure but slightly evolve from one exposure to the next. These aberrations are the dominant limiting term for space-based instruments as well as for ground-based telescopes with state-of-the-art AO systems. Coronagraphs on segmented apertures are now multiplying both on space- and ground-based telescopes. Segment errors in phasing (piston and tip-tilt) introduce specific quasi-static errors before the coronagraph, which can be predicted \cite{leboulleux2018_PairbasedAnalyticalModel} and mitigated
\cite{laginja2021_AnalyticalTolerancingSegmented}. But even continuous mirror telescopes use optics in their instruments which introduce quasi-static aberrations, generally much smaller (associated optical path difference of a few tens of nanometers at most) than the atmospheric aberrations (a few hundreds nanometers). Considering small aberrations, meaning the associated optical path difference negligible compared to the propagation wavelength, Equations~\ref{eq:ephiminus1} and~\ref{eq:intensity_corono} can be written:
\begin{equation}
\begin{split} 
  E_{D,\,\mathrm{S}}\left(\phi_\mathbb{C} \right) &\simeq \mathcal{C}\left[ P\,\left( a\iup + i\,\phi\iup \right) \right]\\
    I_{D,\,\mathrm{S}}\left(\phi_\mathbb{C} \right) &\simeq \left| \mathcal{C}\left[ P\,\left( a\iup + i\,\phi\iup \right)\right]\right|^2
    \label{eq:ED_smallphase}
\end{split}
\end{equation}

\begin{figure}[!ht]
    \centering
    \includegraphics[width=0.6\textwidth]{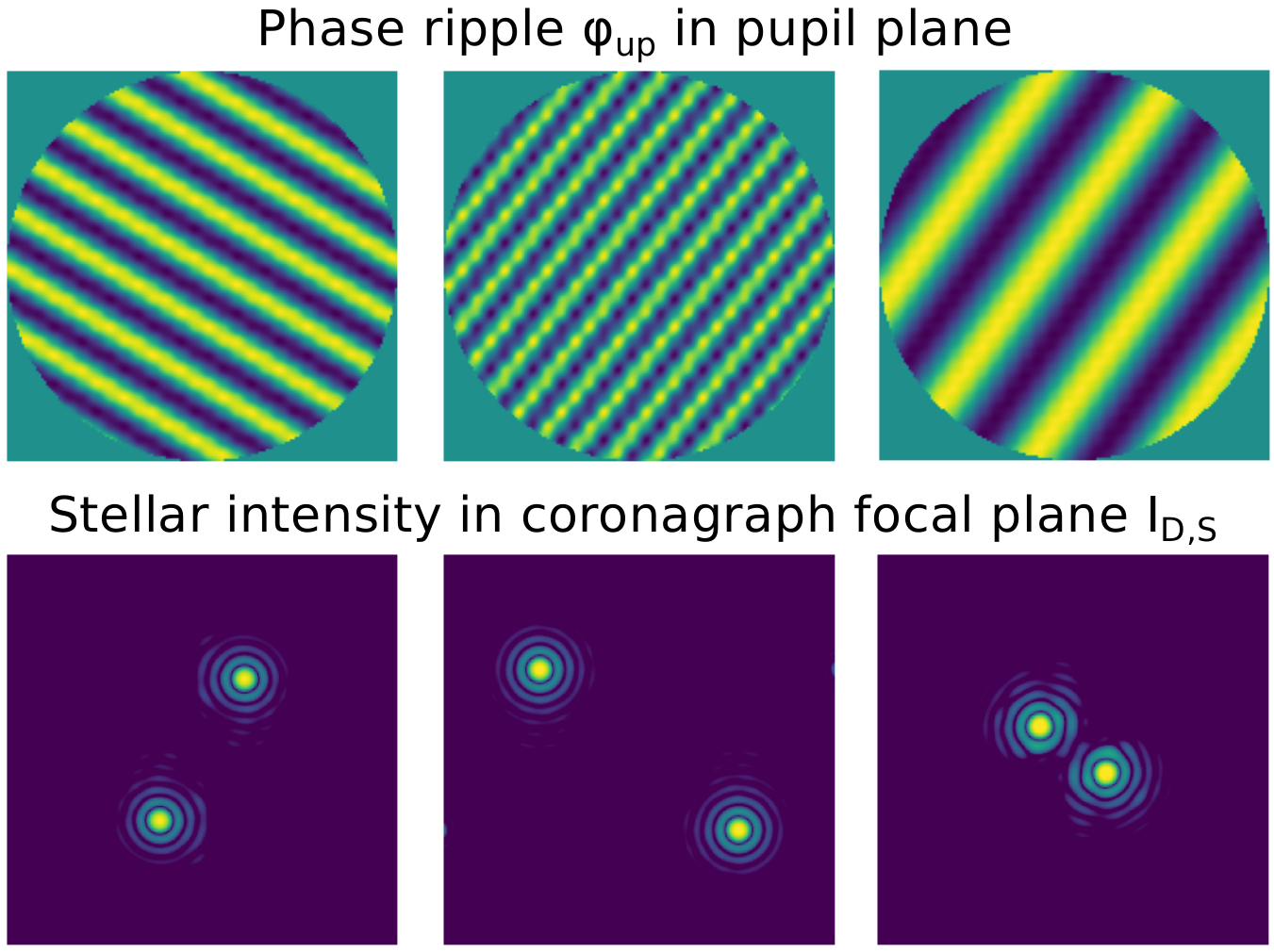}
    \caption{\it \textbf{Top:} Sine phase $\phi\iup$ for three spatial frequencies and directions in the pupil plane. \textbf{Bottom:} Associated coronagraphic image~$I_{D,\,\mathrm{S}}$.}
    \label{fig:sincos_noaberr}
\end{figure}

To understand the creation of speckles in the focal plane~D, we study the effect of a phase only sine shape aberration in the pupil plane. The sine optical path difference ripple has an amplitude~$\sigma\iup$ and frequency~$n/D$ along the horizontal axis of coordinates~$\xi_h$ with~$D$ the pupil diameter:
\begin{equation}
\label{eq:sin_ripple}
\phi\iup(\xi_h) =  \dfrac{2 \pi}{\lambda} \sigma\iup\sin\left(2\pi \frac{n}{D}\,\xi_h\right) 
\end{equation}
If this is the only aberration in the system, and assuming it is small~($ \sigma\iup  \ll \lambda$), we can derive the electric field in the final focal plane from Equations~\ref{eq:C_operator2} and~\ref{eq:ED_smallphase}:
\begin{equation}
E_{D,\,\mathrm{S}} \propto \,E_{0,\lambda}\, \mathrm{FT}\left[L_\lambda\right]\star\left(M_\lambda\,\mathrm{FT}\left[A_\lambda \, P\,i\phi\iup\right]\right)
\label{eq:ed_sin_first}
\end{equation}
The Fourier transform on the right-hand side of the previous equation can be written as:
\begin{equation}
\FT{A_\lambda\,P\,\phi\iup}{\vec{x}} = \,\dfrac{\pi \sigma\iup}{\lambda} \left(\delta\left[\vec{x}+n\,\frac{\lambda\,f}{D}\vec{u}_h\right]-\delta \left[\vec{x}-n\,\frac{\lambda\,f}{D}\vec{u}_h\right] \right) \star \FT{A_\lambda\,P}{\vec{x}}
\end{equation}
with~$\vec{u}_h$ the horizontal unit vector and $\delta$ the~2D~Dirac delta function. We assume that the frequency of the ripple is high enough so that the terms in the previous equation are not modified by the focal plane mask~$M$ (which generally has the most impact close to the center) so that $\mathrm{FT}[P\,\phi\iup]\,M \sim \mathrm{FT}[P\,\phi\iup]$. In this case, Equation~\ref{eq:ed_sin_first} becomes:
\begin{equation}
\label{eq:corono_sine}
E_{D,\,\mathrm{S}}\left(\vec{x}\right) \simeq E_{0,\lambda}\,\dfrac{ \pi \sigma\iup}{\lambda} \left(\delta\left[\vec{x}+n\,\frac{\lambda\,f}{D}\vec{u}_h\right]-\delta \left[\vec{x}-n\,\frac{\lambda\,f}{D}\vec{u}_h\right] \right) \star \FT{A_\lambda\,P\,L_\lambda}{\vec{x}}
\end{equation}
The coronagraphic image for a sine phase function is the sum of two~PSFs located at~$\pm n$ times the resolution element of the instrument (i.e. two speckles at~$\pm n\,\lambda\,f/D$) and with an intensity equals to~$\pi^2\,\sigma\iup^2/\lambda^2$ times the one of the stellar PSF with no focal plane mask~$M_\lambda$.  Figure~\ref{fig:sincos_noaberr} shows examples of different phase ripples creating symmetrical speckles in the intensity focal plane~D. A similar equation for a cosine ripple can be obtained:
\begin{equation}
\label{eq:corono_cosine}
E_{D,\,\mathrm{S}}\left(\vec{x}\right)  \simeq E_{0,\lambda}\,\dfrac{i \pi \sigma\iup}{\lambda} \left(\delta\left[\vec{x}+n\,\frac{\lambda\,f}{D}\vec{u}_h\right]+\delta \left[\vec{x}-n\,\frac{\lambda\,f}{D}\vec{u}_h\right] \right) \star \FT{A_\lambda\,P\,L_\lambda}{\vec{x}}
\end{equation}
These formulas generalize for any direction of the sine and cosine that always induce two symmetrical speckles in the direction of variation of the phase. A more realistic phase is normally composed of a continuous set of spatial frequencies, which can be decomposed in a Fourier series of sines and cosines of individual frequencies of different amplitudes. In the focal plane of the coronagraph, this decomposition results in the speckle field, with individual speckles of different intensities and focal plane locations, shown in Figure~\ref{fig:PSF_coroIm_aberration} (right).

Equation~\ref{eq:corono_sine} shows that the coronagraphic focal plane star light normalized intensity scales with the square of the aberrations. An increase of a factor~2 of the aberration amplitude in the pupil plane increases the intensity in the focal plane by a factor~4. The maximum of the coronagraphic intensity normalized by the maximum of the~PSF recorded with no coronagraph is roughly~$(\pi\,\sigma\iup/\lambda)^2$. Table~\ref{tab:contrast_vs_rms} shows the expected normalized intensity of the individual speckles in the focal plane for different levels of the pupil sine phase aberration.
\begin{table}[!ht]
{\setlength{\extrarowheight}{3pt}
\begin{tabular}{|c|c c c c|}
\hline
Optical path difference amplitude~$\sigma\iup$ &100\,nm&1\,nm&10\,pm&1\,pm\\
\hline
Starlight speckle normalized intensity&$1$&$10^{-4}$&$10^{-8}$&$10^{-10}$\\
\hline
\end{tabular}}
\vspace{0.3cm}
\caption{\it Approximate normalized intensity for a given level of sin/cos optical path difference.}
\label{tab:contrast_vs_rms}
\vspace{-0.6cm}
\end{table}

To detect an Earth-like planet~$10^{10}$ times fainter than its star the aberrations must be of the order of~$1\,$pm at the spatial frequency that creates the speckle at the planet's position. A realistic quasi-static phase is composed of a continuous set of spatial frequencies though (see speckle field in Figure~\ref{fig:PSF_coroIm_aberration}, right). It can be shown that~$10^{-10}$ level can be reached in the coronagraph image with~$\sim0.1\,$nm~rms phase aberrations over the pupil. The exact value depends on the power spectral density of the aberrations but in any case, such a small value cannot be obtained by construction. Active optical elements can be used to minimize the speckle intensity and reach the equivalent of picometric aberrations at given spatial frequencies~(Sections~\ref{sec:wfc} to~\ref{sec:experimentalvalidation}).

Very similar equations and reasoning could be obtained starting with an amplitude ripple:
\begin{equation}
\label{eq:sin_ripple_amplitude}
a\iup(\xi) = a_0\sin\left(2\pi \frac{n}{D}\,\xi_h\right) 
\end{equation}
We do not develop these equations in this paper, but phase and amplitude play a similar role in the creation of the speckle field.

\section{Measuring the performance of a coronagraphic system}
\label{sec:good_coro}

There is hardly a consensus in the community on the exact definition and terminology of the metrics that can be used to measure the performance of a coronagraphic instrument. In any case, the first goal of a coronagraphic system is to detect an exoplanet with a given $F_{\mathrm{P}}/F_{\mathrm{S}}$ flux ratio at a given separation~$\vec{x}$ on the detector. The first two criteria are therefore the attenuation of the starlight~(Section~\ref{subsec:normalized_intensity}) and the transmission of the exoplanet signal~(Section~\ref{subsec:exoplanet_thpt}), which strongly impact the SNR of the detection (Section~\ref{subsec:snr_corono}). However, other parameters need to be considered when designing an instrument for a given science case, such as the spectral bandwidth and a team must make compromises between different metrics to achieve the scientific objectives on real telescopes~(Section~\ref{subsec:real_coronagraphs}).

\subsection{Normalized intensity}
\label{subsec:normalized_intensity}
The first goal of the coronagraph is to minimize the star's intensity $I_{D,\,\mathrm{S}}$ resulting from the electric field in Equation~\ref{eq:ephiminus1}. The most common metrics used to measure the efficiency of this minimization at a given point $\vec{x}$ in the focal plane is the stellar normalized intensity $\eta_{\mathrm{S}}$:
\begin{equation}
   \eta_{\mathrm{S}}\left(\vec{x}\right) = \displaystyle \frac{\int_\mathcal{A}I_{D,\,\mathrm{S}}\left(\vec{x} +\vec{u}\right) \mathrm{d}\vec{u}}{\int_\mathcal{A} \mathrm{PSF}\left(\vec{u}\right) \mathrm{d}\vec{u} }
   \label{eq:eta_s}
\end{equation}
where $\mathcal{A}$ is the region of interest in the focal plane. This region of interest can be an aperture of the size of the telescope resolution element (Figure~\ref{fig:airy}, left) or a single detector pixel. We choose a single pixel area which subsequently leads to: 
\begin{equation}
   I_{D,\,\mathrm{S}}\left(\vec{x}\right) = \eta_{\mathrm{S}}\left(\vec{x}\right) \mathrm{PSF}(0) = \eta_{\mathrm{S}}\left(\vec{x}\right) \mathrm{max} \left( \mathrm{PSF} \right)
\end{equation}
This quantity is sometimes referred as "raw contrast" in publications, but because there is no consensus in the community on the definition of "contrast", we purposefully decided to avoid using this term in this review and we refer to it as "normalized intensity" instead.

Normalized intensity values usually range from $1$ in the center of the focal plane in the absence of a coronagraph to lower than $10^{-10}$ or better for very good coronagraphic systems. 2D-images showing the coronagraphic normalized focal plane images are often used to study the performance, as seen in Figure~\ref{fig:PSF_coroIm_aberration}. A~1D-radial profile of the normalized intensity as a function of the angular separation from the star is also often plotted. The profile is usually calculated using an azimuthal average (usually called mean normalized intensity) or azimuthal standard deviation (usually called~$1\sigma$ normalized intensity) accounting or not for statistical biases~\cite{mawet2014_small_statistics}. Finally, the performance can be expressed with a single number, the average or standard deviation of the normalized intensity within a given region of the coronagraphic image (for example the region where the speckle intensity is minimized, see Section~\ref{subsec:corr_1DM}). 

\subsection{Exoplanet throughput and inner working angle}
\label{subsec:exoplanet_thpt}

The only way to probe a coronagraph's performance is the measurement of the signal-to-noise ratio of the  exoplanet signal in the stellar speckle field. This ratio is therefore not only dependent on the amount of residual starlight in the coronagraphic focal plane (normalized intensity), but also on the amount of light from the the exoplanet that reaches the focal plane. This is often called the coronagraph's "planetary throughput". We define the planetary throughput $\eta_{\mathrm{P}}$ as:
\begin{equation}
  \eta_{\mathrm{P}}\left(\vec{x}\right)=\displaystyle\frac{ \int_\mathcal{A}I_{D,\,\mathrm{P}}\left(\vec{x} +\vec{u}\right) \mathrm{d}\vec{u}}{\int_\mathcal{A} \mathrm{PSF}\left(\vec{u}\right) \mathrm{d}\vec{u} }
   \label{eq:throughput_def}
\end{equation}
with $\mathcal{A}$ a given area in the focal image. The most used region of interest is an aperture of radius~$0.7 \, \lambda/D$ (Full width Half max of an Airy pattern, the PSF for a clear round aperture, see Figure~\ref{fig:airy}). This definition takes into account the transmission of the instrument (quantity of planetary light that goes through the instrument and reaches the focal plane), but also how the off-axis image shape is distorted by the instrument. 

Because we normalize by the telescope's PSF with no coronagraph, this definition does not take into account the effect of the telescope aperture itself on the planetary signal. In some cases, it is necessary to compare two different instruments on two different telescopes with different apertures (e.g. an off-axis telescope and an on-axis telescope). In this case, the PSF of the instrument in Equation~\ref{eq:throughput_def} can be replace by the PSF created by a clear round aperture of the same size. 

\begin{figure}[!ht]
    \centering
    \includegraphics[width=\textwidth]{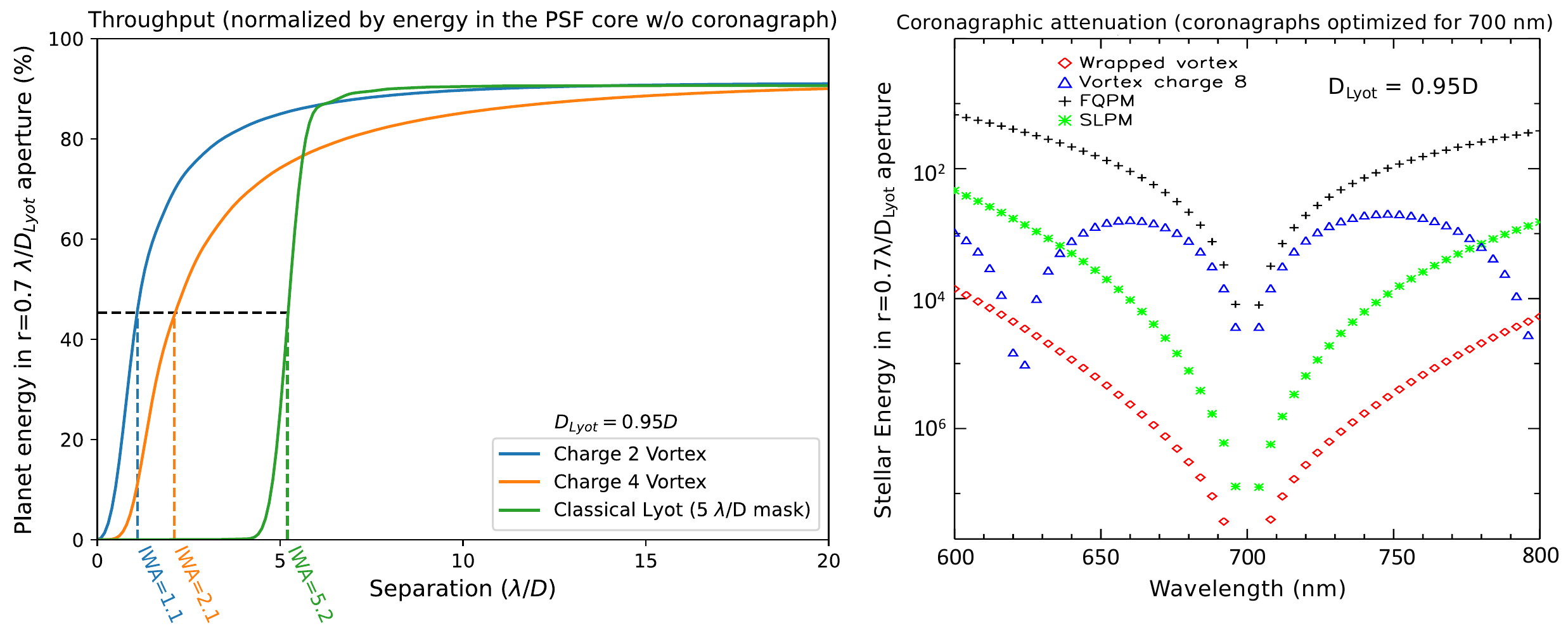}

    \caption{\it {\bf Left}: Numerically simulated throughput for classical Lyot, vortex charge~2 and~4 coronagraphs~\cite{mawet2005_AnnularGroovePhase}. Vertical dashed lines give the~IWA for each coronagraph. {\bf Right} (adapted from \cite{galicher2020_FamilyPhaseMasks}): Numerical simulations of monochromatic performance as a function of wavelength for wrapped vortex \cite{galicher2020_FamilyPhaseMasks}, charge 8 vortex \cite{mawet2005_AnnularGroovePhase}, four-quadrant phase mask \cite{Rouan00} and six level phase mask \cite{Hou2014Wide-band} coronagraphs optimized at~$700\,$nm.}
    \label{fig:performances}
\end{figure}

Because coronagraphic systems are designed to remove all on-axis light, throughput inevitably tends towards zero at small angular separations and usually increases with the distance from the star. The throughput usually ranges from almost 100\% for coronagraphs with clear apertures to a few percents or even less for apertures with central obscurations and discontinuities. One can show the impact on the off-axis~image~(Equation~\ref{eq:intensity_corono}) by plotting the maximum of the planet intensity at every point of the field of view. Because historically, the throughput of most coronagraphs is solely a radial function only, it is often plotted as a function of the separation to the star. Figure~\ref{fig:performances}~(left) shows such a plot for three coronagraphs. For the classical Lyot coronagraph, the throughput is very close to~0 below~$5\,\lambda/D$ where the light is blocked by the focal plane mask. As soon as the separation is larger than the focal plane mask radius~($5\,\lambda/D$), the transmission is almost~$100\,\%$. For the vortex phase mask coronagraphs, the transmission gradually increases from~0 to its maximum, which enables the detection of exoplanets at smaller separations.

We often summarize such a curve using the inner working angle~(IWA) of the coronagraph. This number represents the maximum separation at which the coronagraph throughput is halved:
\begin{equation}
     \eta_{\mathrm{P}}\left(IWA\right) = \dfrac{\mathrm{Max}(\eta_{\mathrm{P}})}{2}
   \label{eq:iwa}
\end{equation}
The~IWA are indicated by dashed vertical lines in Figure~\ref{fig:performances}. The IWA needs to be used carefully because it does not encompass the complexity of throughput curves. For example, the throughput is almost a discontinuous function for Lyot coronagraphs (the exoplanet is either behind the mask or not) whereas they gradually increase with separation for vortex coronagraphs. 

\subsection{Signal-to-noise ratio of a coronagraphic instrument}
\label{subsec:snr_corono}

Now that we have introduced the metrics that define the effects of a coronagraphic system on the star's on-axis light (normalized intensity) and on the planetary off-axis light (throughput), the calculus of the SNR of a high-contrast detection introduced in Section~\ref{subsec:why_corono} can be revisited. Aberrations induce "unknown" speckles that can mimic the exoplanet's image in the focal plane of the coronagraph. The noise variance now includes the speckle intensity, as well as the photon noise of both the speckles and the exoplanet. The signal-to-noise ratio of the exoplanet detection~(Equation~\ref{eq:init_SNR}) can be written as:
\begin{equation}
\mathrm{SNR}_{P|S}(T_{exp}) = \frac{F_{\mathrm{P}} \eta_{\mathrm{P}}\left(\vec{x}\right) T_{exp} \mathrm{PSF}(0) }{\sqrt{F_{\mathrm{S}}^2 \eta_{\mathrm{S}}^2\left(\vec{x}\right) T_{exp}^2 \mathrm{PSF}^2(0)  +F_{\mathrm{P}} \eta_{\mathrm{P}}\left(\vec{x}\right) T_{exp} \mathrm{PSF}(0) + F_{\mathrm{S}} \eta_{\mathrm{S}}\left(\vec{x}\right) T_{exp} \mathrm{PSF}(0) }}
\label{eq:snr_speckle_and_photon_noise}
\end{equation}
If speckles are much brighter than the exoplanet, the equation becomes
\begin{equation}
\mathrm{SNR}_{P|S}(T_{exp}) \simeq \,\frac{F_{\mathrm{P}}}{F_{\mathrm{S}}}\frac{\eta_{\mathrm{P}}\left(\vec{x}\right)}{\eta_{\mathrm{S}}\left(\vec{x}\right) }
\end{equation}
In this case, the~SNR does not depend on the exposure time. No matter how long the exposure, the SNR is set by the exoplanet to star flux ratio, the planet throughput~$\eta_{\mathrm{P}}$ and the normalized intensity~$\eta_{\mathrm{S}}$ that depends on the aberration level (Equations~\ref{eq:ED_smallphase} and~\ref{eq:eta_s}). In this case, speckle intensity must necessarily be minimized using active correction (Section~\ref{sec:wfc}) or post-processing (Sections~\ref{sec:postproc} to~\ref{sec:experimentalvalidation}) to detect the exoplanet's signal.

If the exoplanet is much brighter than the speckles, Equation~\ref{eq:snr_speckle_and_photon_noise} can be written as:
\begin{equation}
\mathrm{SNR}_{P|S}(T_{exp}) \simeq \sqrt{F_{\mathrm{P}} \eta_{\mathrm{P}}\left(\vec{x}\right) T_{exp} \mathrm{PSF}(0)}=\sqrt{\eta_{\mathrm{P}}}\,\mathrm{SNR}_{P}(T_{exp}) 
\end{equation}
with~$\mathrm{SNR}_{P} (T_{exp})$ the photon noise~SNR of the exoplanet in the absence of star. In this case, the SNR no longer depends on the speckle intensity anymore. This shows the importance of designing coronagraphs that minimize the on-axis starlight~(Equation~\ref{eq:coro_on_axis}) but also maximize the off-axis exoplanet throughput~$\eta_{\mathrm{P}}$~(Equation~\ref{eq:coro_off_axis}), as shown in \cite{mazoyer2018_ActiveCorrectionApertureII, ruane2016_ApodizedVortexCoronagraph}. This is especially true for coronagraphic systems aimed at high-contrast levels with complex apertures (large central obscurations and/or spiders): the gain in normalized intensity is often reached at great cost for the off-axis~image shape and therefore produces low coronagraph throughput \cite{carlotti2013_ApodizedPhaseMask,guyon2014_HighPerformanceLyot, ndiaye2016_ApodizedPupilLyot, fogarty2017_PolynomialApodizersCentrally, pueyo2013_HighcontrastImagingArbitrary, trauger2016_HybridLyotCoronagraph, zimmerman2016_ShapedPupilLyot, mazoyer2018_ActiveCorrectionApertureI, ruane2016_ApodizedVortexCoronagraph}. 

To be sure to encompass all the effects when benchmarking the coronagraph designs, HabEx~\cite{gaudi2020_HabitableExoplanetObservatory} and LUVOIR~\cite{theluvoirteam2019_LUVOIRMissionConcept} teams combined the normalized intensity and throughput at each point of the focal plane~D to estimate the yield of exoplanets detected by these coronagraphic systems out of hundreds of Monte Carlo draws of possible companions around neighboring stars \cite{stark2019_ExoEarthYieldLandscape}. This technique can be quite time consuming but allows a thorough comparison between different coronagraphic systems.

\subsection{Designing coronagraphs for real instruments}
\label{subsec:real_coronagraphs}
An important parameter to evaluate the performance of a coronagraph is the spectral bandwidth. This parameter is usually measured in~$\%$ ($\Delta \lambda / \lambda$). Some coronagraphic systems are very chromatic~\cite{Roddier97,riaud2001_FourQuadrantPhaseMaskCoronagraph} whereas others have been designed to be used with large bandwidths~\cite{soummer2003_dzpm, galicher2011_MultistageFourquadrantPhase,ndiaye2012_ImprovedAchromatizationPhase, delorme2016_FocalPlaneWavefront, delorme2016_LaboratoryValidationDualzone, cady2017, galicher2020_FamilyPhaseMasks}. One way to show this performance is to plot the attenuation of the on-axis star at several wavelengths (Figure~\ref{fig:performances}, right). Usually, systems are optimized for a given wavelength (here~700\,nm) and their performance degrades as the difference with the optimal wavelength increases. When designing an instrument further studies are required to understand the impact of the bandwidth: change in the images with wavelength (and not only the central attenuation), wavefront control performance~(sections~\ref{sec:wfc} to~\ref{sec:experimentalvalidation}) over large bandwidths, exoplanet throughput, etc.

Coronagraphic systems are designed using optimized solutions which usually first attempt to predict the performance for an ideal, static, telescope (no aberrations and perfectly stable). However, some designs are also optimized to resist against to unknown but expected variations, like low-order aberrations (tip, tilt, defocus, astigmatism). Indeed, the robustness of a coronagraph system to these aberrations is important to predict expected performances in real conditions. Robustness is usually inversely correlated with the~IWA: small~IWA coronagraphs are very sensitive to the centering of the star~image onto the focal plane mask~(tip-tilt). The amount of low-order aberrations expected on the telescope (ground- or space-based) therefore strongly impacts the design of the coronagraphic system and in turn influences the~IWA. This parameter is usually more important for ground-based telescopes than for space-based telescopes that are more stable.

Finally, the simplicity of one's coronagraphic design must be considered. For example, some teams suggested coronagraphs with using multiple optics in cascade and agreed that keeping the alignment stable can be very challenging~\cite{galicher2011_MultistageFourquadrantPhase,mawet2013_double_vortex}. The coronagraphs must be simple enough so that the alignment is relatively fast and stable over long periods of time.

To conclude, the "best universal coronagraph" does not exist. Depending on the telescope (complexity of the aperture, expected aberrations and stability, capabilities of the AO system), the type of exoplanets targeted (warm Jupiters, exo-Jupiters or exo-Earths), type of host star (magnitude, spectrum) and the type of analysis involved (imager, spectrograph), some coronagraphs perform better than others. Hence, several metrics should be used in parallel to optimize the coronagraph instrument, while always being driven by the science case. 

\section{Wavefront Control}
\label{sec:wfc}

\subsection{Objective: minimization of the speckle intensity}
\label{subsec:wfc_goal}
In Section~\ref{sec:coro_aberration}, we showed that phase and amplitude aberrations strongly limit the coronagraph's performance~(Equation~\ref{eq:ED_smallphase} and Table~\ref{tab:contrast_vs_rms}). To compensate for quasi-static phase and amplitude aberrations, teams have long suggested the use of active optical devices like deformable mirrors~\cite[DM]{malbet1995_HighDynamicRangeImagingUsing, borde2006_HighContrastImagingSpace}) or spatial light modulators~\cite[SLM]{kuhn2017_ImplementingFocalplanePhasea, kuhn2021_SLMbasedActiveFocalplane}). Such devices are used to introduce pure phase shifts~$\phi\DM$ on the beam they reflect or transmit. SLMs are enticing by the high number of correction frequencies they offer. However, they are usually liquid-crystal based: as such they are quite chromatic and work in polarized light. That is why the overwhelming majority of high-contrast testbeds and instruments currently rely on~DMs for optical control. In the remainder of this paper, we shall focus on the~DM case. A review of existing~DM technology in the context of astronomy can be found in \cite{madec2012_OverviewDeformableMirror}. Contrarily to~AO, high-contrast imaging correction does not require high speed or important stroke~DMs (quasi-static aberrations are small and vary slowly) but need to access high spatial frequencies (i.e. require a large number of actuators). 

\begin{figure}[!ht]
    \centering
    \includegraphics[width=\textwidth]{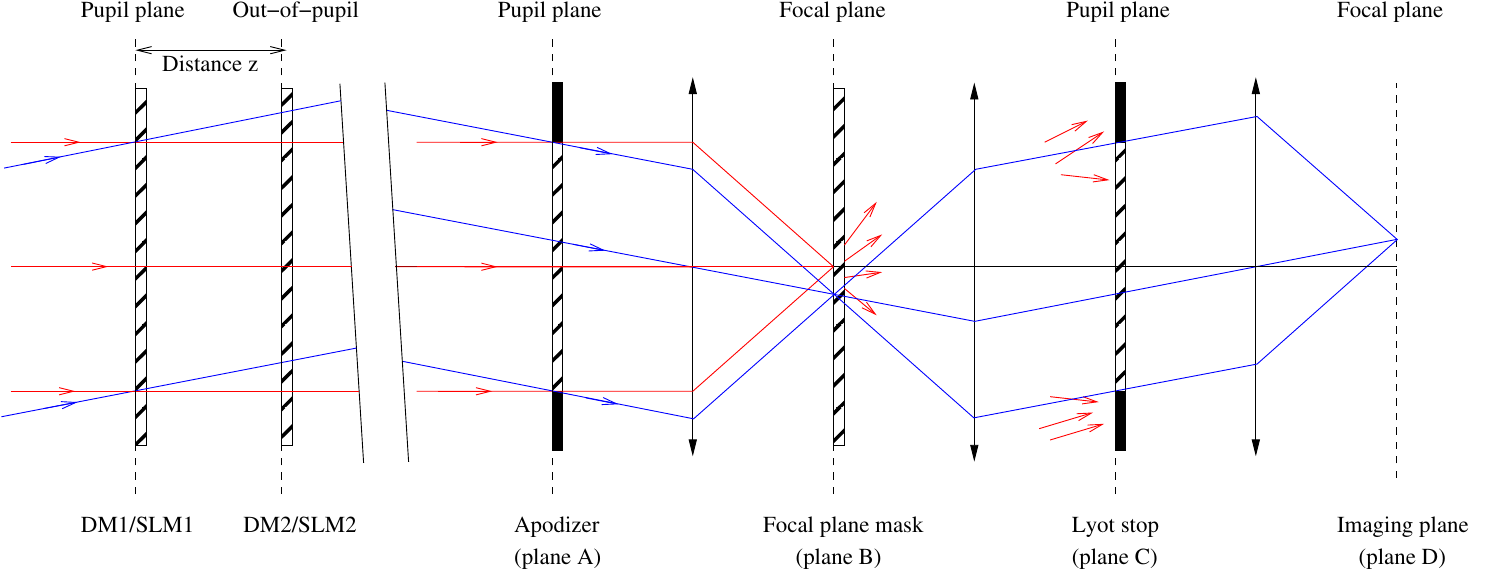}
    \caption{\it Stellar coronagraph associated to two active optical corrector (deformable mirror,~DM, or spatial light modulator,~SLM). See Figure~\ref{fig:principle_coro} and text for details.}
    \label{fig:principle_coro_DM}
\end{figure}

As shown in Figure.~\ref{fig:principle_coro_DM}, they can be set in or out of a pupil plane upstream the pupil plane~A. Assuming~$\mathcal{C}[P]=0$, small aberrations and a~DM conjugated to the pupil plane, Equation~\ref{eq:ED_smallphase} that gives the electric field in the imaging plane~D becomes:
\begin{equation}
  E_{D,\,\mathrm{S}} \simeq \mathcal{C}\left[ P\,\left( a\iup + i\,\left[\phi\iup+\phi\DM\right] \right) \right]
   \label{eq:Ed_phiDM}
\end{equation}
The goal of the correction is to find a specific phase~$\phi\DM$ (i.e voltages) that minimizes the stellar speckle intensity inside a given area called dark hole~\cite{malbet1995_HighDynamicRangeImagingUsing}. An important aspect to note is that this is different from the goal of classical~AO systems discussed in Section~\ref{subsec:adaptive_opt}, which minimizes the total phase in the pupil plane. Equation~\ref{eq:Ed_phiDM} shows that, in the case of no amplitude aberrations~($a\iup=0$), choosing $\phi\DM=-\phi\iup$ cancels the star's intensity in the imaging plane. This is however not possible because $\phi\iup$ is composed of an infinite number of spatial frequencies whereas $\phi\DM$ is not (DM has a finite number of actuators). Other solutions for~$\phi\DM$ are therefore favored to minimize~$I_{D,\,\mathrm{S}}$ as explained hereafter.

\subsection{Half dark hole using one pupil plane deformable mirror}
\label{subsec:corr_1DM}
\begin{figure}[!ht]
    \centering
    \includegraphics[width=0.6\textwidth]{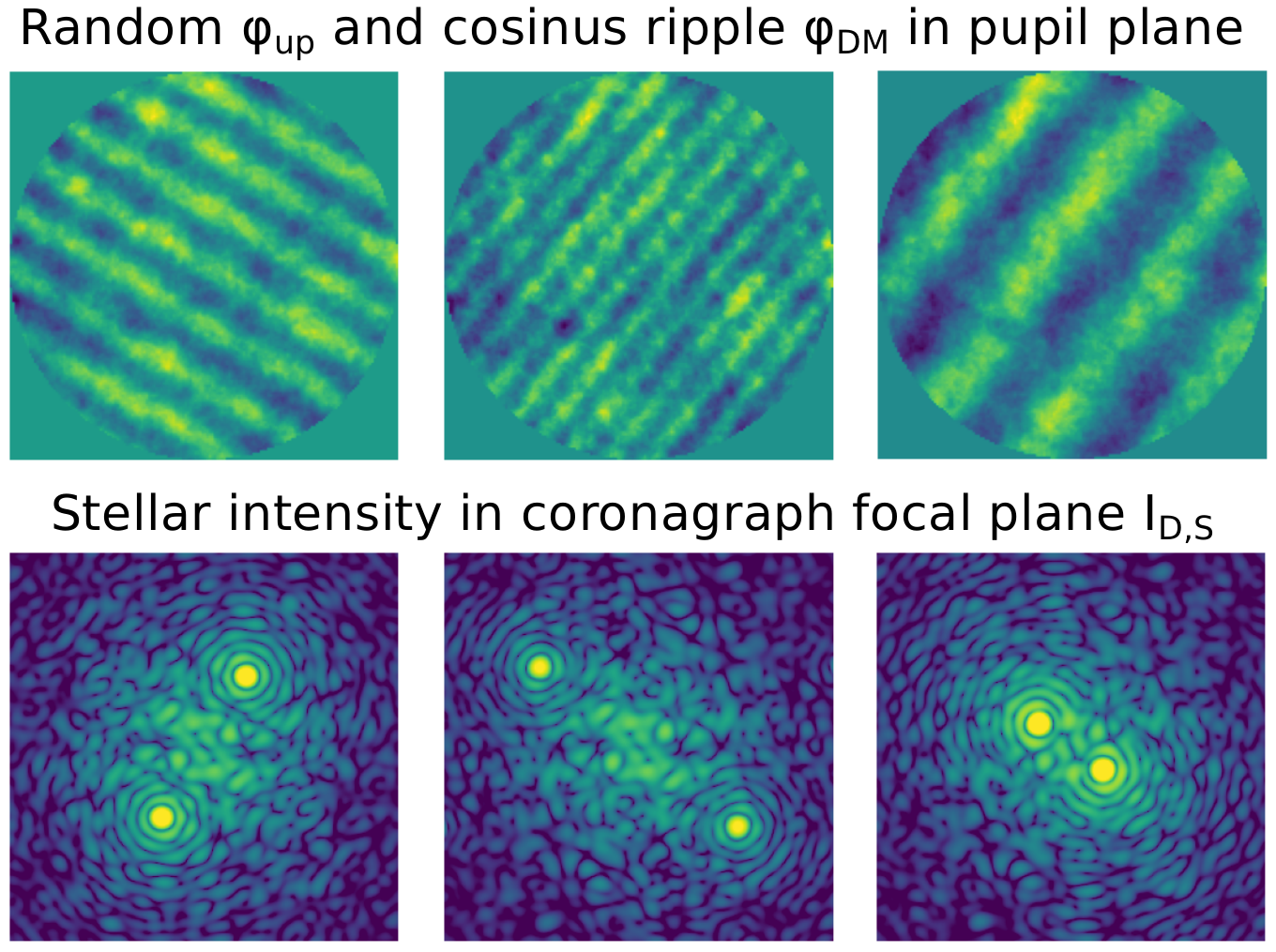}
    \caption{\it \textbf{Top:} Phase resulting from the sum of a random aberrations~$\phi\iup$ and one cosine~$\phi\DM$ applied on the~DM for three frequencies and directions. \textbf{Bottom:} Associated coronagraphic images~$I_{D,\,\mathrm{S}}$.}
    \label{fig:sincos_withaberr}
\end{figure}

To understand how a~DM conjugated to the pupil plane impacts the focal plane of the coronagraph, consider a single phase ripple on this~DM with a phase amplitude of~$2\,\pi\,\sigma_{DM}/\lambda$. Following the formalism already used in Section~\ref{subsec:quasi_static}:
\begin{equation}
\label{eq:sin_ripple_DM}
\phi_{DM}(\xi_h) =  \dfrac{2 \pi}{\lambda} \sigma_{DM}\sin\left(2\pi \frac{n}{D}\,\xi_h\right) 
\end{equation}
The resulting phase $\phi\iup+\phi\DM$ is shown on the first row of Figure~\ref{fig:sincos_withaberr} for three different frequencies. Using the same reasoning as in Section~\ref{subsec:quasi_static}, we can show that the phase ripples~$\phi_{DM}$ on the~DM modifies two speckles of the existing speckle field created by~$\phi\iup$ (Figure~\ref{fig:sincos_withaberr}, second row).
By decomposing any~DM achievable phase as a sum of sines and cosines (Fourier series decomposition), we obtain a basis of~DM functions with very local responses in the coronagraphic image~\cite{poyneer2005_OptimalModalFouriertransform}. 

The smallest spatial period introduced by the~DM is limited by the number of actuators, and it is equal to~$N_{\mathrm{act}}/2$ with~$N_{\mathrm{act}}$ the number of actuators across the pupil diameter in the considered direction. In this case, applying the Nyquist–Shannon sampling theorem~(see Section~2.4.1 in \cite{goodman2005_IntroductionFourierOptics}) the two speckles are at the maximum distance from the optical axis:
\begin{equation}
\label{eq:numax}
x_{h,\mathrm{max}}=\dfrac{N_{\mathrm{act}}}{2}\,\frac{\lambda\,f}{D}
\end{equation}
Hence, the~DM can modify the stellar intensity within a finite area of the coronagraphic image~$I_D$, the influence zone, going from the optical axis to~$\pm N_{\mathrm{act}}\,\lambda\,f/(2\,D)$. A~DM with more actuators in the pupil results in a larger influence zone.

Dark holes can be chosen with different shapes and sizes, so long as they are inside the influence zone~\cite{malbet1995_HighDynamicRangeImagingUsing}. They are often chosen smaller than the full~DM influence zone and limiting the correction to fewer frequencies often improves the starlight minimization~\cite{mazoyer2013_EstimationCorrectionWavefront}. Most authors refer to the dark hole inner and outer edges as the inner working angle ($IWA_{DH}$) and outer working angle ($OWA_{DH}$). These definitions must not be confused with the coronagraph~IWA introduced in Section~\ref{subsec:exoplanet_thpt}. 

\begin{figure}[!ht]
    \centering
    \includegraphics[width=0.8\textwidth]{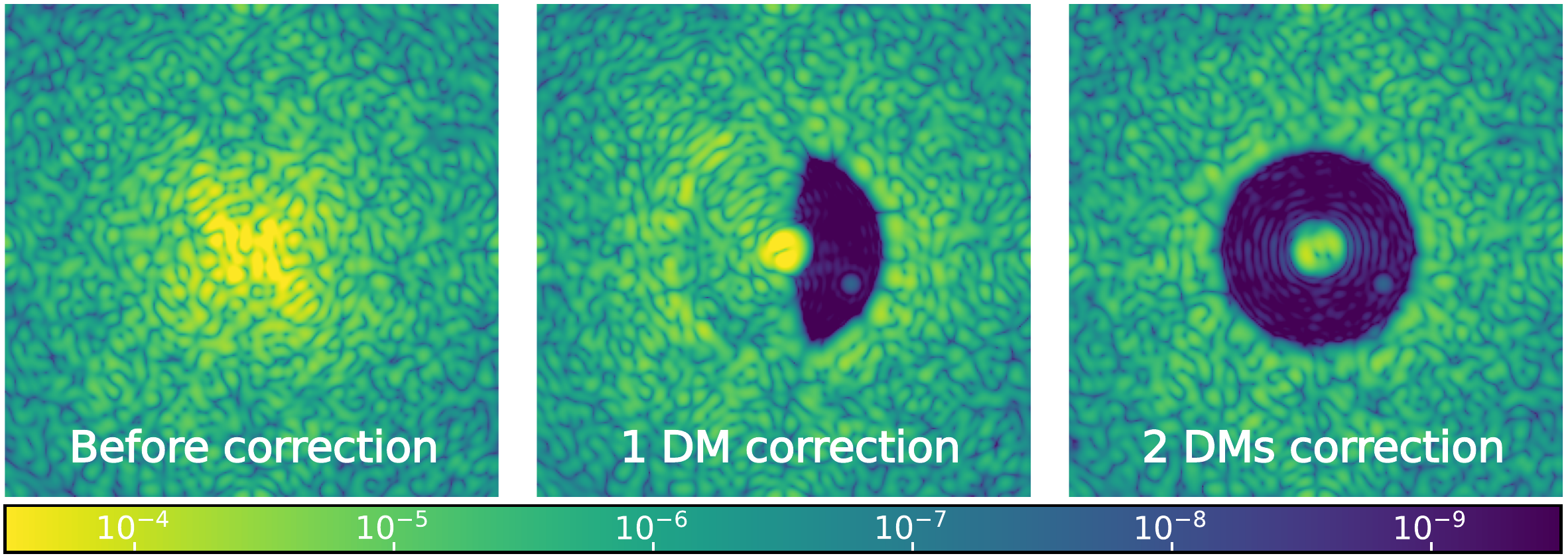}
    \caption{\it Numerical simulations of the coronagraphic image using a four-quadrant phase mask in the presence of small static aberrations for a narrow bandwidth, before correction (left), after~1\,DM correction (center) or~2\,DM correction (right). A Jupiter-like planet is located at~$7\,\lambda/D$ on the bottom right and it is $3.10^{-8}$ fainter than the star. The aperture is fully circular. DMs have~$32\times32$ actuators and they are controlled using~\cite{mazoyer2018_ActiveCorrectionApertureI} assuming a perfect estimation of the electric field in the coronagraphic image. Dark holes go from~$3\,\lambda/D$ to $10\,\lambda/D$. The color bar represents the normalized intensity.}
    \label{fig:dark_holes}
\end{figure}

Equation~\ref{eq:Ed_phiDM} shows that speckles from phase and amplitude aberrations $a\iup + i\,\phi\iup$ are to be minimized. DMs can only introduce pure phase aberrations $i\,\phi\DM$ in the plane they are located into, and they cannot expect to correct both phase and amplitude at the same time inside a dark hole centered on the optical axis. They can however correct for both in half the focal plane. This capability comes from the property that the Fourier transform of a pure imaginary function~($i\,\phi\DM$ in the pupil plane) exhibits an anti-Hermitian symmetry (anti-symmetric real part and symmetric imaginary part). Using the small phase assumption, Equation~\ref{eq:ed_sin_first} links the electric field in the focal plane to the Fourier transform of $i\,\phi\DM$. This shows that whatever the phase introduced by the~DM, the induced electric field on one side of the final focal plane totally determines the field on the other side by anti-Hermitian symmetry. However, the combination of phase and amplitude aberrations $a\iup + i\,\phi\iup$ is neither real nor purely imaginary, resulting in a speckle field with no clear symmetry. The~DM can therefore only minimize one side of the speckle field, creating a half-dark hole correction. Figure~\ref{fig:dark_holes} (left) shows a numerical simulation of a speckle field before any correction. An exoplanet is located at~$7\,\lambda/D$ from the star but is undetectable because it is $3.10^{-8}$ times fainter than the star. Figure~\ref{fig:dark_holes} (center) shows the minimization of the speckle intensity inside a half dark hole from~$3\,\lambda/D$ to~$10\,\lambda/D$ using one deformable mirror. Inside the dark hole, the speckle intensity is minimized down to a few~$10^{-9}$ of the maximum of the star~PSF recorded with no coronagraph, allowing the planet detection, on the bottom right. Note that a half-dark hole could have been created in any directions and could have a different shape, as long as it is inside a one half plane and only requires spatial frequencies smaller than the maximum distance the~DM can reach~(Equation~\ref{eq:numax}).

\subsection{Full dark hole using two deformable mirrors in cascade}
\label{subsec:corr_2DM}
We now use two~DMs sequentially to compensate for both phase~$\phi\iup$ and amplitude~$a\iup$ aberrations in the pupil plane:~DM$_1$ is set in a pupil plane and~DM$_2$ outside of the pupil plane, at a distance~$z$, as shown in Figure~\ref{fig:principle_coro_DM}. Because~DM$_2$ is outside of the pupil plane, the Fraunhofer approximation is not longer valid. The more general Fresnel approximation (Section~4.2. in \cite{goodman2005_IntroductionFourierOptics}) can be used to describe the propagation of the WFS in this plane. However, for the specific case of a sine (or cosine) ripple, the Fresnel formalism can be simplified using the Talbot effect (Section~4.5.2. in \cite{goodman2005_IntroductionFourierOptics}). This is a diffraction effect describing the propagation of a pure sine aberration, shifting from pure phase to phase and amplitude as the propagating distance increases, before shifting back to phase at a distance called the Talbot length~$z_T=2\,D^2/(n^2\,\lambda)$. A pure small sine phase~(Equation~\ref{eq:sin_ripple_DM}) introduced by~DM$_2$ converts into phase~$\phi_\mathrm{DM_2}$ and amplitude~$a_\mathrm{DM_2}$ in the pupil plane~\cite{zhou2010_talbot}:
\begin{equation}
\begin{split} 
\phi_{\mathrm{DM}_2} &= \dfrac{2 \pi}{\lambda} \sigma_{\mathrm{DM}_2}\sin\left(2\pi \frac{n}{D}\,\xi_h\right)\,\cos\left(2\,\pi\,\frac{z}{z_T}\right)\\
a_{\mathrm{DM}_2} &= \dfrac{2 \pi}{\lambda} \sigma_{\mathrm{DM}_2}\sin\left(2\pi \frac{n}{D}\,\xi_h\right)\,\sin\left(2\,\pi\,\frac{z}{z_T}\right)
\end{split}
\end{equation}
These equations can also be expressed as a function of the Fresnel number of the system~$\mathcal{F}=D^2/(\lambda\,z)$:
\begin{equation}
\begin{split} 
\phi_{\mathrm{DM}_2} &= \dfrac{2 \pi}{\lambda} \sigma_{\mathrm{DM}_2}\sin\left(2\pi \frac{n}{D}\,\xi_h\right)\,\cos\left(\pi\,\frac{n^2}{\mathcal{F}}\right)\\
a_{\mathrm{DM}_2} &= \dfrac{2 \pi}{\lambda} \sigma_{\mathrm{DM}_2}\sin\left(2\pi \frac{n}{D}\,\xi_h\right)\,\sin\left(\pi\,\frac{n^2}{\mathcal{F}}\right)
\end{split}
\label{eq:2dm_fresnel}
\end{equation}
Hence, a phase ripple introduced by the second~DM with a spatial frequency $n/D$ such as $n=\sqrt{\mathcal{F}/2}$ is converted into a pure amplitude ripple~$a_{\mathrm{DM}_2}$ in the pupil plane. It can then be used to compensate for one amplitude aberration~$a\iup$ at this spatial frequency. DM$_1$ is then used to compensate for the phase aberrations~$\phi\iup$.

However, for a different spatial frequency or wavelength,~DM$_2$ introduces both phase~$\phi_{\mathrm{DM}_2}$ and amplitude~$a_{\mathrm{DM}_2}$ in the pupil plane. It can still be used to compensate for the amplitude aberration~$a\iup$ but now~DM$_1$ must compensate for both the phase aberrations~$\phi\iup$ and the additional phase~$\phi_{\mathrm{DM}_2}$ introduced by~DM$_2$. Therefore, there is no distance~$z$ that enables the use of~DM$_2$ for perfect amplitude correction at all spatial frequencies and all wavelengths. The position~$z$ can be optimized considering the size of the chosen dark hole, the number of actuators, the level of phase and amplitude aberrations to be corrected and the bandwidth of observation. Several authors have analyzed in detail these dependencies to find the optimal position of the~DMs~\cite{shaklan2006_ReflectivityOpticalSurface, Pueyo_2007_Polychromatic_Compensation, beaulieu2017_HighcontrastImagingSmall, beaulieu2020_HighContrastSmall, mazoyer2017_FundamentalLimitsHighcontrast}. Normalized intensity limits for the one or two~DM cases are recalled in~\cite{pueyo2018_DirectImagingDetection}. Using two~DMs, speckles can now be corrected in a ~$360^\circ$ dark hole, as shown on Figure~\ref{fig:dark_holes} (right). 

\subsection{Need for a model of the instrument}
\label{subsec:wfc_techniques}
Once we know how~DMs impact the electric field in the coronagraphic focal plane image, one can wonder what the best shape to be applied on them is to minimize the stellar speckle intensity. Several control methods have been developed for focal plane wavefront control in the case of one or several~DMs correction. These have been particularly well reviewed and explained in Groff et al. (2016)~\cite{groff2016_MethodsLimitationsFocal} or more recently in French by Potier (2020) \cite{potier20phd}. One can cite speckle nulling~\cite{borde2006_HighContrastImagingSpace}, electric field conjugation~\cite{giveon2007_broadbandefc} or stroke minimization~\cite{pueyo2009_OptimalDarkHole}.

Most of the techniques use a model of the light propagation inside the instrument. An interaction matrix (also called Jacobian) is built to link the effect of each~DM actuator voltage to the electric field in the coronagraphic image. The matrix can be fully computed based on the optical model or directly recorded with the instrument~\cite{galicher2010_SelfcoherentCameraFocal}.
In both cases, the interaction matrix is then inverted to obtain the control matrix (also called command matrix). Hence, once the electric field in the final focal plane~D is measured~($E_{D,\,\mathrm{S}}$, section~\ref{sec:fpwfs}), it is multiplied by the command matrix to obtain the voltages to be sent to the~DMs in order to minimize the speckle intensity inside the dark hole. The matrix inversion usually uses singular value decomposition-based techniques \cite{boyer1990_AdaptiveOpticsInteraction} with different forms of truncation of the singular values and by adding other constraints (e.g. reaching a given lower speckle intensity with the minimum value of~DM voltages possible).

All current focal plane wavefront control techniques work in closed loop because nor the estimation of~$E_{D,\,\mathrm{S}}$~(Section~\ref{sec:fpwfs}) nor the controller are perfect. For example, the use of an interaction matrix to model the impact of the~DM voltages in the focal plane relies on the assumption that the coronagraph system is linear with aberrations. This relies on the assumption of small aberrations (if two aberrations are summed, their effects in the final plane~D are linearly added, see Equation~\ref{eq:ED_smallphase}) and the linearity of the~DM (if two voltages maps are summed on the DM, the result is the sum of the two phases). Some techniques are exploring non-linear solutions~\cite{paul2013_CoronagraphicPhaseDiversity,herscovici-schiller2018_ExperimentalValidationNonlinear}. They use more accurate but also more complex theoretical models that rely on more parameters that need to be calibrated. A simple model is easy to calibrate on real instruments but returns slightly approximated voltages and inefficient or slow corrections. A trade-off must be found between complex models that takes too long to calibrate and simple models that may be inefficient.

\section{Focal Plane Wavefront Sensing} 
\label{sec:fpwfs}

\subsection{Phase degeneracy}
\label{subsec:phase_degeneracy}
The closed loop described in Section~\ref{subsec:wfc_techniques} requires the knowledge of the electric field~$E_{D,\,\mathrm{S}}$ in the focal plane~D. As explained in Section~\ref{subsec:quasi_static}, adaptive optics systems use a dedicated channel for measuring and minimizing the phase~$\phi\iup$ but they cannot measure the electric field in the focal plane~D. Hence, the field~$E_{D,\,\mathrm{S}}$ needs to be directly measured from the coronagraphic image using a focal plane wavefront sensor~(FPWFS). The electric field of the speckle in Equation\,\ref{eq:Ed_phiDM} becomes:
\begin{equation}
    E_{D,\,\mathrm{S}}=|E_{D,\,\mathrm{S}}|\exp{\left(i\, \mathrm{Arg}\left[ E_{D,\,\mathrm{S}} \right]  \right)}
\end{equation}
The detector in the focal plane can only measure the light's intensity, which is given by: 
\begin{equation}
    I_{D,\,\mathrm{S}}=|E_{D,\,\mathrm{S}}|^2
\end{equation}
All information about the argument of the complex field $\mathrm{Arg}\left[ E_{D,\,\mathrm{S}} \right]$ is lost. This is known as the phase degeneracy problem. A given speckle field intensity can be created by an infinity of different electric fields. However, the minimization of the speckle intensity by the control algorithm requires a precise estimation of the electric field. This degeneracy is the main challenge of the~FPWFS. 

The problem of phase degeneracy has long  been studied in adaptive optics or microscopy~\cite{Fienup1982_Phase_retrieval_comparison}. However, these algorithms cannot be directly applied to coronagraphs.

\subsection{Focal plane wavefront sensing: modulation of the speckle intensity}
Two recent reviews describe focal plane wavefront sensing techniques in details~\cite{groff2016_MethodsLimitationsFocal, jovanovic2018_ReviewHighcontrastImaging}. In this section, we therefore only briefly recall the context.

Many methods have been suggested to break the degeneracy described in Section~\ref{subsec:phase_degeneracy}, relying on the incoherence between the light waves coming from the star and from its environment (exoplanets, circumstellar disks, another star, etc). These approaches can be separated into two major categories: spatial coherent modulation and temporal coherent modulation. 
The former uses part of the starlight rejected by the coronagraph to create one interference pattern in the coronagraphic image so that the stellar speckle intensity is spatially modulated unlike the exoplanet image. Among these, the self-coherent camera uses a modified Lyot stop~\cite{baudoz2006_SelfCoherentCameraNew, galicher2010_SelfcoherentCameraFocal}, the coronagraphic modal wavefront sensor uses a specific phase mask in the pupil plane to sense low-order modes~\cite{Wilby2017_CMWs} and the kernel phase~WFS uses an asymmetry of the pupil~\cite{Martinache2013AsymmetricPupil}. 
In the temporal modulation, several coronagraphic images are recorded applying different known phase aberrations to the beam so that the speckle intensity is modulated from one image to the others. Among these, the pair-wise probing technique~\cite{borde2006_HighContrastImagingSpace, giveon2007_ClosedLoopDM,giveon2011_PairwiseDeformableMirror,potier2020_ComparingFocalPlane} introduces various known phases using the~DM (called probes), this is probably the most used~FPWFS currently. The COFFEE method introduces a defocus or actuator pokes with the~DM~\cite{paul2013_CoronagraphicPhaseDiversity,herscovici-schiller2018_ExperimentalValidationNonlinear} and finally, the slow but robust speckle nulling technique modulates the intensity of individual speckles using the~DM~\cite{bottom2016_SpeckleNullingWavefronta,martinache2014_OnSkySpeckleNulling}.

The temporal modulation techniques are attractive because they require no modification of the instrument's configuration, as they use the available~DMs to produce the modulation. They need a sequence of several coronagraphic images for one estimation of~$E_{D,\,\mathrm{S}}$. Hence, their performance is degraded if the quasi-static aberrations~(and the associated speckles) change between the recorded images. Finally, the calibration can last longer than the observing time dedicated to record astrophysical signal. Spatial modulation techniques require a specific instrumental configuration. This can be a drawback for the implementation of the technique in an existing instrument. Only one image is required to measure~$E_{D,\,\mathrm{S}}$, which is much more efficiently than temporal modulation techniques. Another important advantage of spatial modulation is that they do not degrade the normalized intensity performance during measurements. They can therefore be used during the science sequence. An experimental comparison of spatial and temporal modulation techniques showed that similar performance were reached under space conditions in laboratory~\cite{potier2020_compare_WFS}

Finally, some techniques have been developed to stabilize the speckle pattern. They measure small phase differences with respect to reference focal plane image. These techniques cannot create a dark hole, but they are used to maintain the normalized intensity during the observation, inside an already achieved dark hole. Among them, we can cite the linear dark field control technique~\cite{guyon2017_Spectralldfc,miller2017_SpatialLinearDark,currie2020_LaboratoryDemonstrationSpatial}, techniques to control low order aberrations~\cite{mas2012_TiptiltEstimationCorrection,huby2015_PostcoronagraphicTiptiltSensing} and others~\cite{Pogorelyuk2019Dark_Hole_Maintenance}.

As for correction techniques~(Section~\ref{subsec:wfc_techniques}), most of the~FPWFS rely on a model of light propagation in the coronagraph instrument to estimate the electric field of the speckle~$E_{D,\,\mathrm{S}}$ from a known temporal or spatial modulation. For example, the small aberration assumption is often used. As for controllers, simple models may produce an approximate estimate requiring numerous iterations in the correction loop to create the dark hole. More elaborate models might produce more accurate estimates but they require regular calibrations. Again, a trade-off needs to be found. Finally, several authors have recently tried to use neural networks trained on instrument data to approach the instrumental models in order to produce faster algorithms \cite{sun2018_NeuralNetworkControl,quesnel2020_DeepLearningbasedFocal}.

All these~FPWFS only sense light intensity that is modulated. Unmodulated light that reaches the focal plane can be from astrophysical sources (exoplanet or circumstellar disk). This property can be used in post-processing for coherent differential imaging~(section~\ref{sec:postproc}). Unfortunately, some of the starlight itself present in the focal plane can also be unmodulated. This can happen in the presence of different polarization states, diffusion by the optics or because of speckles varying faster than the integration time and averaging into a halo~(Section~\ref{subsec:adaptive_opt}). Some authors call the unmodulated stellar light the incoherent light~\cite{pogorelyuk2021_InformationtheoreticalLimitsRecursive}.

\begin{figure}[!ht]
    \centering 
    \includegraphics[width=.9\textwidth]{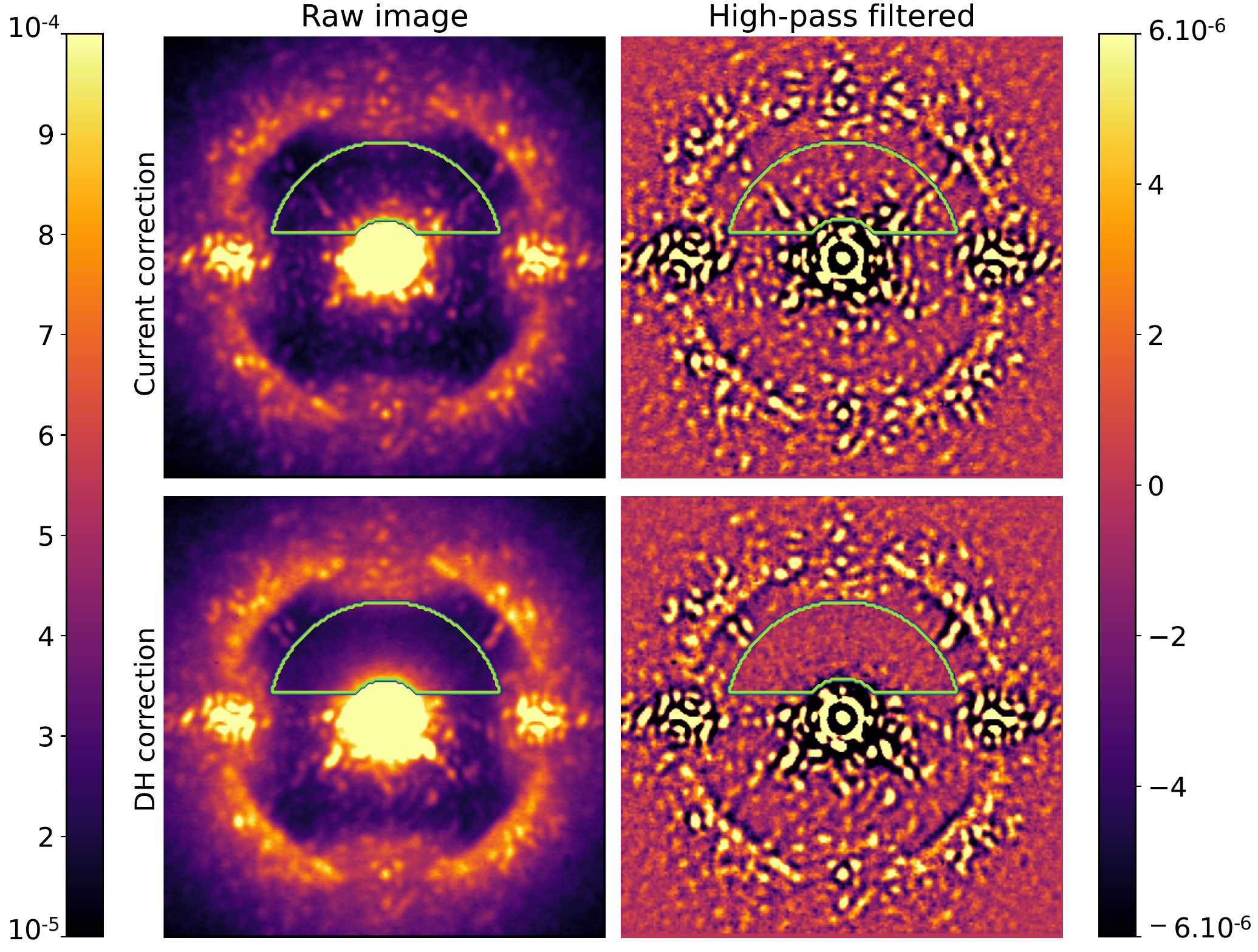}
    \caption{\it On-sky SPHERE/IRDIS coronagraphic images after AO correction and, before (\textbf{top}) and after (\textbf{bottom}) speckle minimization inside the dark hole (green line), using pair-wise probing and electric field conjugation \cite{potier2020_IncreasingRawContrast}. {\bf Left:} Raw images. The central part that is saturated is a smooth halo produced by~AO residuals. Quasi-static speckles are also visible. {\bf Right:} After high-pass filtering to show the quasi-static speckles only. The color bar represents the normalized intensity. Adapted from Potier et al. (2022) \cite{Potier2022_onsky_dh}.}
    \label{fig:potier_dh_sphere}
\end{figure}

\section{From laboratory to on-sky validation}
\label{sec:experimentalvalidation}
There are numerous technical solutions for each stage of the coronagraphic instruments: apodizer, focal plane mask, Lyot stop, focal plane wavefront sensor, low order aberration wavefront sensor, wavefront controller, etc. And the coupling of several of these is required to build an efficient coronagraphic instrument. There is however no perfect combination that enables the detection of any faint source around any bright source in any observing conditions. To understand the pros and cons of each combination, technical demonstrations are required in laboratory and/or on the internal laser source of existing instruments, until their validation on real instruments, usually during dedicated technical time.

Several optical testbeds are currently used to validate individual techniques or different optical configurations. The existing high-contrast testbeds and recent results are described in~\cite{mazoyer2019_HighContrastTestbedsFuture}.
These testbeds can be in air \cite{potier2020_compare_WFS, Llop-Sayson2020_HCST, soummer2021_hicat, currie2020_LaboratoryDemonstrationSpatial} or in a vacuum \cite{Riggs2021_supernyquist, Marx2021_spie}. Once the techniques have been validated in the laboratory, certain instruments already installed on the telescope offer the possibility of technical validation using internal laser sources during daylight: SCEXAO at Subaru Telescope \cite{bos2019_FocalplaneWavefrontSensing}, SPHERE at Very Large Telescope \cite{potier2020_IncreasingRawContrast}, MagAO at Magellan Telescope \cite{miller2019_SpatialLinearDark}, KPIC at Keck Observatory~\cite{mawet2016_KPIC}.
Finally, the last step of validation is done on-sky. We can cite on-sky demonstrations at Palomar Observatory~\cite{Galicher2019_scc_palomar}, Subaru Telescope~\cite{Bos2021_ldfc_onsky_scexao}, Keck Observatory~\cite{bos_2021_fast_furious} or Very Large Telescope~\cite{vigan2022_CalibrationQuasistaticAberrationsa}.

One example of a recent on-sky demonstration is shown in figure~\ref{fig:potier_dh_sphere}. This is the first on-sky efficient minimization of the quasi-static speckle intensity inside a dark hole. The demonstration was done at the VLT on the SPHERE/IRDIS detector observing the bright star HIP~57013, using the SAXO~\cite{sauvage2010_SAXOEXtremeAdaptive} adaptive optics (AO) system with an apodized pupil lyot coronagraph and a focal plane mask of~$185\,$mas in radius. The focal plane wavefront sensing is achieved using the pair-wise probing method the DM are controlled with the electric field conjugation technique~\cite{potier2020_ComparingFocalPlane, potier2020_IncreasingRawContrast}. The images show only stellar light: no off-axis source (planet or disk) has been detected around HIP\,57013~($I_{D,\,\mathrm{S}}$, Equation~\ref{eq:ephiminus1}) so far. In the two images on the left, the central part that is saturated is a smooth halo with decreasing intensity from the center. It results from the aberrations that are too fast to be corrected by the~AO system~(Section~\ref{subsec:adaptive_opt}). The quasi-static speckles that are detected are induced by phase and amplitude aberrations~(Section~\ref{subsec:quasi_static}) in the telescope and instrument. The circular darker zone is the~AO controlled region, corresponding to spatial frequencies in the pupil plane at which the~AO system minimizes the phase aberrations~$\phi\iup$ (Equation~\ref{eq:numax}). The images on the right are the same as on the left but they have been high-pass filtered to uniquely show the speckles. The top images are the starting point once the SAXO~AO loop is closed. The bottom images are the results after the focal plane wavefront control loop is closed to minimize the stellar intensity inside a dark hole~(green line). A half dark hole was used because SPHERE has a single~DM~(Section~\ref{subsec:corr_1DM}). All speckles are removed from the dark hole leaving the smooth halo of the~AO residuals only~(incoherent light), which cannot be corrected because it varies faster than the exposure time. Inside the dark hole, the normalized intensity is improved by a factor of~$\sim6$ to reach~$\sim5.10^{-6}$ at~200\,mas and~$\sim 10^{-6}$ between~$300$ and~$660\,$mas in the high-pass filtered images. The gain depends on the observing conditions, the time dedicated to the correction loop and the exposure time (here the detector noise is the main limitation inside the dark hole). To remove the speckles in the bottom part of these images (outside the dark hole), the authors propose the use of coherent differential imaging (section~\ref{sec:postproc}).

\section{Post-processing of coronagraphic images: differential imaging}
\label{sec:postproc}

The images from Figure~\ref{fig:potier_dh_sphere} are a good example of what is obtained with current instruments~(top left) and in the near future~(bottom left). In both cases, part of the coronagraphic image is dominated by stellar speckles that mimic exoplanet images and eventually a smooth halo of starlight. These can easily mask the astrophysical signal~(exoplanets and circumstellar disks). The halo adds photon noise to the~SNR of the exoplanet detection. Its impact can be reduced by integrating over longer time periods. Another solution is the upgrade of the adaptive optics system for a faster correction (and a reduced halo intensity). For convenience, hereafter, the term "speckles" refers to any type of starlight residual that reaches the detector, AO residual halo or quasi-static speckles. Several imaging techniques and observing strategies have been used to enhance the detection capabilities of coronagraphic instruments. 

The reader can find more information on this subject in these two recent reviews~\cite{pueyo2018_DirectImagingDetection,currie2022_review_imdiff}. In the present review, we quickly recall the principle and limitations of these techniques~(Section~\ref{subsec:goal_post-proc}) and, we present some of the specific observational strategies (Section~\ref{subsec:imdif_strategies}) and post-processing algorithms (Section~\ref{subsec:imdif_dataproc}).

\subsection{Goal of differential imaging post-processing techniques}
\label{subsec:goal_post-proc}
If we use a single raw coronagraphic image (for example the one shown on the top left of Figure~\ref{fig:potier_dh_sphere}), it is not possible to differentiate a speckle from an exoplanet. However, the behavior of speckles and exoplanet images differ in specific ways. Differential imaging techniques exploit such differences to calibrate the speckles and extract the astrophysical signal. In an ideal case, the speckle pattern is perfectly removed leaving the astrophysical signal and photon noise. The~SNR of the exoplanet detection~(Equation~\ref{eq:snr_speckle_and_photon_noise}) then can be written as:
\begin{equation}
\mathrm{SNR}_{P|S}(T_{exp}) = \frac{F_{\mathrm{P}} \eta_{\mathrm{P}}\left(\vec{x}\right) T_{exp} \mathrm{PSF}(0) }{\sqrt{F_{\mathrm{P}} \eta_{\mathrm{P}}\left(\vec{x}\right) T_{exp} \mathrm{PSF}(0) + F_{\mathrm{S}} \eta_{\mathrm{S}}\left(\vec{x}\right) T_{exp} \mathrm{PSF}(0) }}
\end{equation}
For very bright stars, differential imaging can enhance the normalized intensity of the raw coronagraphic images by a factor of~100 at best~\cite{langlois2021_sphere_perf}. However, photon noise cannot be subtracted, and this is the final theoretical limit of these techniques. Using the same reasoning as in Section~\ref{subsec:why_corono}, we can use the SNR of the planet if it were observed alone, without a coronagraph $\mathrm{SNR}_{P} (T_{exp}) = \sqrt{F_{\mathrm{P}} \mathrm{PSF}(0) T_{exp}}$:
\begin{equation}
    \mathrm{SNR}_{P|S}(T_{exp}) = \mathrm{SNR}_{P}(T_{exp}) \sqrt{\eta_{\mathrm{P}}\left(\vec{x}\right)} \left( 1+ \frac{ F_{\mathrm{S}} \eta_{\mathrm{S}}\left(\vec{x}\right)  }{F_{\mathrm{P}} \eta_{\mathrm{P}}\left(\vec{x}\right)} \right)^{-1/2}
\end{equation}
This shows that the post-processing techniques can increase the detection SNR but the theoretical limit is still very dependent on the active minimization of the speckle intensity~($\eta_{\mathrm{S}}$).

\subsection{Strategies of observation for differential imaging}
\label{subsec:imdif_strategies}
\begin{figure}[!ht]
    \centering
    \includegraphics[width=.5\textwidth]{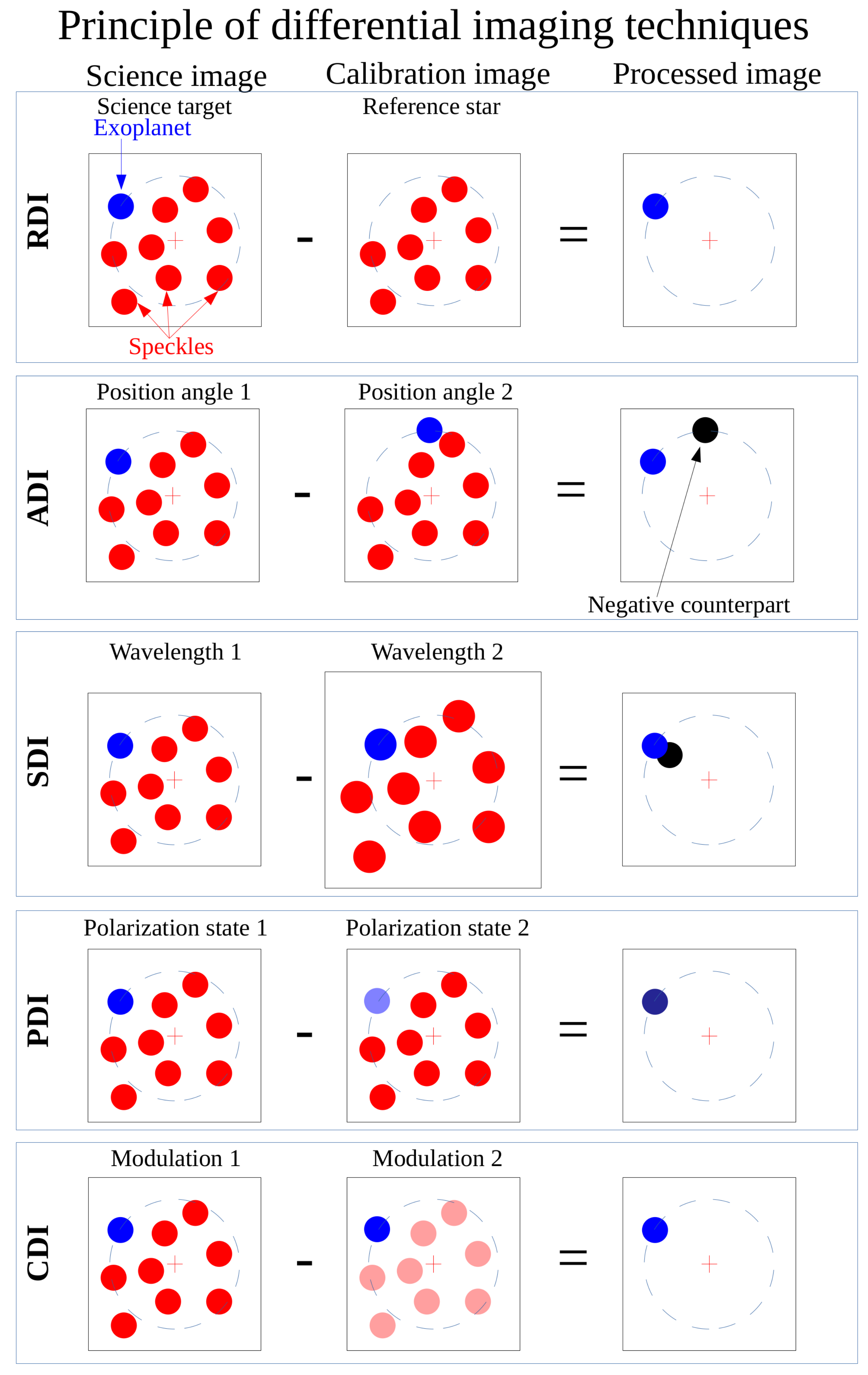}
    \caption{\it Diagram of differential imaging techniques (RDI, ADI, SDI, PDI, CDI) assuming a perfect calibration of the speckle pattern. Left: science image composed of the exoplanet image to be detected (blue) and the stellar speckle pattern (red). Center: image used for calibrating the speckle pattern. Right: Subtraction of the calibrated image from the science image. This "subtraction" may include other operations like spatial scaling for~SDI, rotation of the field-of-view for ADI, etc. Black represents over-subtraction of exoplanet signal.}
    \label{fig:im_diff_principle}
\end{figure}
Strategies of observation have been used for years to record sets of images so that differential imaging can be applied to calibrate the stellar speckle pattern and extract the exoplanet signal: angular differential imaging~\cite[ADI]{marois2006_AngularDifferentialImaging}, dual-band imaging and spectral differential imaging~\cite[SDI]{rosenthal1996_SDI,racine1999_SDI,marois2004_sdi,thatte2007_sdi}, reference differential imaging~\cite[RDI]{beuzit1997_RDI,choquet2016_RDI,ren2021_RDI,wahhaj2021_RDI}, polarization differential imaging~\cite[PDI]{baba2003_MethodImageExtrasolar,baba2005_PDI}, and coherent differential imaging~\cite[CDI]{codona2004_CDI,baudoz2006_SelfCoherentCameraNew,bottom2017_CDI,gerard2018_fast_CDI,bos2021_PESCC}. All of these strategies aim at modulating the speckle pattern and the exoplanet signal differently. Figure~\ref{fig:im_diff_principle} shows a simplified diagram of a science image (left column), a calibration image (center) and a processed image after differential imaging (right). The stellar speckle pattern is in red and the image of the planet in blue. The central cross is the optical axis on which the image of the star is centered. In reality, more than two images are used for both the science image and the calibration image, but the principle remains the same.

Each strategy relies on specific assumptions about the speckle pattern. Using~RDI, similar stars are observed under similar instrumental set-up assuming the speckle pattern is stable over time. Using~ADI, one assumes that most of the optical aberrations remain static during observations and, come from planes that are optically conjugated with the pupil plane. Keeping the pupil orientation fixed, the speckle pattern is stable in the images whereas the field-of-view rotates around the central star. Using dual-band imaging, the spectrum of the star (and equally, of the speckles) is supposed to be different from the exoplanet's spectrum. Using~SDI, one assumes the speckles are induced by achromatic optical path differences in a pupil plane so that the evolution of the speckle intensity with wavelength is known. Using~PDI, one considers that unlike the starlight, the exoplanet light is partially polarized. Finally, using~CDI one assumes that the speckle pattern is stable over time for temporal modulation of the speckle intensity~\cite{codona2004_CDI,bottom2017_CDI}. No assumptions are needed when using spatial modulation of the speckle intensity~\cite{baudoz2006_SelfCoherentCameraNew,baudoz2013_LaboratoryTestsPlanet,gerard2018_fast_CDI,bos2021_PESCC}. For a more complete description of each strategy, the reader can refer to~\cite{currie2022_review_imdiff}.

\begin{figure}[!ht]
    \centering
    \includegraphics[width=\textwidth]{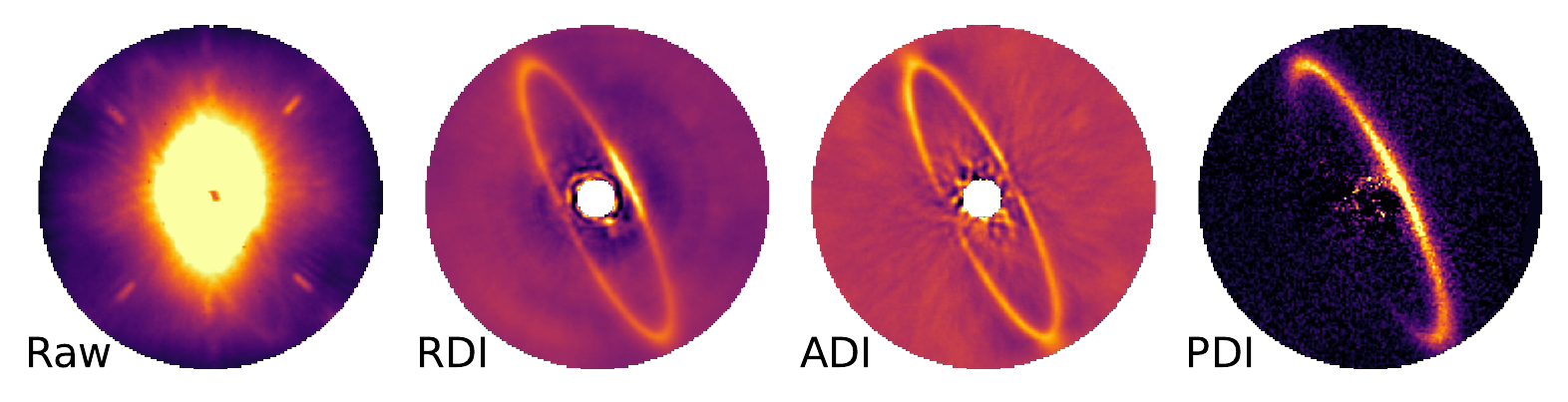}
    \caption{\it  HR 4796 images in H band from GPI data and several differential imaging techniques. North is up and East on the right. {\bf From Left to Right}: Raw image in total intensity, the disk is barely noticeable ; {\bf RDI} leaving the disk almost with no distortion ; {\bf ADI} that usually results in a higher~SNR but the self-subtraction often heavily impacts the disk shape (mostly visible at the ansae and regions close to the star) ; {\bf PDI} that is not impacted by self-subtraction but can only detect regions where the light is most polarized (front part of the disk), while regions with lower polarized flux (back side of the disk) are not detected. Raw/RDI/ADI images are adapted from \cite{chen2020_MultibandGPIImaging}, PDI image from \cite{arriaga2020_MultibandPolarimetricImaging}.}
    \label{fig:im_diff_hr47_gpi}
\end{figure}

Figure~\ref{fig:im_diff_hr47_gpi} shows GPI data of the HD~4796 debris disk. On the left is the raw coronagraphic image~($I_{D,\,\mathrm{S}}$) dominated by the adaptive optics halo and quasi-static speckles. In addiction, the post-processed images using~RDI, ADI and~PDI are shown with the detected belt of dust. We notice that images show different shapes and structures of the disk. Each technique probes one part of the signal (e.g. PDI probes polarized light only). Moreover, each technique introduces biases (e.g. self-subtraction explained hereafter).

\subsection{Several algorithms for differential imaging}
\label{subsec:imdif_dataproc}
For each observation strategy, several post-processing algorithms have been published to extract the exoplanet's signal. Classical~ADI \cite{marois2006_AngularDifferentialImaging}, locally optimized combination of images \cite[LOCI]{lafreniere2007_NewAlgorithmPointspread}, template-LOCI~\cite{marois2014_tloci}, principal component analysis~\cite{soummer2012_PCA,amara2012_PYNPOINTImageProcessing}, subtraction of median images or radial profiles~\cite{galicher2018_AstrometricPhotometricAccuracies}, use of statistical models~\cite{cantalloube2015_andromeda,flasseur2018_paco,flasseur2021_rexpaco}.

All of these algorithms need a cube of several raw coronagraphic images~$I_D$ (focal plane~D in Figure~\ref{fig:principle_coro_DM}) recorded at different instants~$t$, wavelengths~$\lambda$ or polarization states~$p$ (or a combination of them). Using notations defined at the end of Section~\ref{subsec:form_coro_no_aberr}, the recorded intensities~$I_D$ are the sum of the exoplanet's intensity~$I_{D,\,\mathrm{P}}$ and of the stellar speckle pattern~$I_{D,\,\mathrm{S}}$ (Eq.~\ref{eq:ID_star_planet})
\begin{equation}
I_D(\vec{x},\lambda,t,p) = I_{D,\,\mathrm{S}}(\vec{x},\lambda,t,p)+ I_{D,\,\mathrm{P}}(\vec{x},\lambda,t,p)
\end{equation}
For example, in the case of~ADI observations, $I_D$ only depends on~$\vec{x}$ and~$t$ and the orientation of the field of view with respect to the North changes with~$t$.

To extract the exoplanet's signal~$I_{D,\,\mathrm{P}}(\vec{x},\lambda,t,p)$, algorithms usually combines the frames~$I_D$ to derive an estimation~$I^{\mathrm{est}}_{D,\,\mathrm{S}}$ of the speckle pattern~$I_{D,\,\mathrm{S}}$, with the notable exception of~\cite{flasseur2018_paco,flasseur2021_rexpaco}. Details on the calculation of~$I^{\mathrm{est}}_{D,\,\mathrm{S}}$ for each techniques are given in~\cite{galicher2018_AstrometricPhotometricAccuracies}. The algorithms then subtract the estimated speckle patterns from each frame and create a residual datacube~$R$
\begin{equation}
R(\vec{x},\lambda,t,p) =  I_{D,\,\mathrm{S}}(\vec{x},\lambda,t,p) + I_{D,\,\mathrm{P}}(\vec{x},\lambda,t,p)- I^{\mathrm{est}}_{D,\,\mathrm{S}}(\vec{x},\lambda,t,p)
\end{equation}
Finally, all frames of~$R$ are mean- or median-combined to sum up the exoplanet's signal. 

If the calibration of~$I_{D,\,\mathrm{S}}$ is perfect, only the exoplanet's signal remains in~$R$. In reality, the calibration is not perfect because of the principle of the technique (e.g. self-subtraction for~ADI, SDI, PDI) or because the assumptions listed above are not verified (e.g. stability of the speckle pattern). As a result:
\begin{itemize}
    \item $I^{\mathrm{est}}_{D,\,\mathrm{S}}\ne I_{D,\,\mathrm{S}}$;
    \item part of the exoplanet signal is present in~$I^{\mathrm{est}}_{D,\,\mathrm{S}}$. 
    \end{itemize}
Subsequently, in~$R$, part of the exoplanet's signal is removed, and part of the star speckle pattern remains. The former effect is known as self-subtraction and can be partially calibrated for point-like sources such as exoplanets~\cite{pueyo2016_DetectionCharacterizationExoplanets,pueyo2018_DirectImagingDetection,galicher2018_AstrometricPhotometricAccuracies} and less easily for extended sources like circumstellar disks~\cite{milli2012_selfsubtraction_diskADI,esposito2014_ModelingSelfsubtractionAngular,mazoyer2020_DiskFMForwardModeling}. This is why extracting an accurate astrometry and photometry/spectrometry on images processed by differential imaging is very challenging.

Finally, as $I^{\mathrm{est}}_{D,\,\mathrm{S}}\ne I_{D,\,\mathrm{S}}$, differential imaging techniques cannot remove all of the starlight from the coronagraphic image. The difference between~$I^{\mathrm{est}}_{D,\,\mathrm{S}}$ and~$I_{D,\,\mathrm{S}}$ may be smaller with future instruments that are expected to be more stable than current ones.

\begin{figure}[!ht]
    \centering
    \includegraphics[width=.7\textwidth]{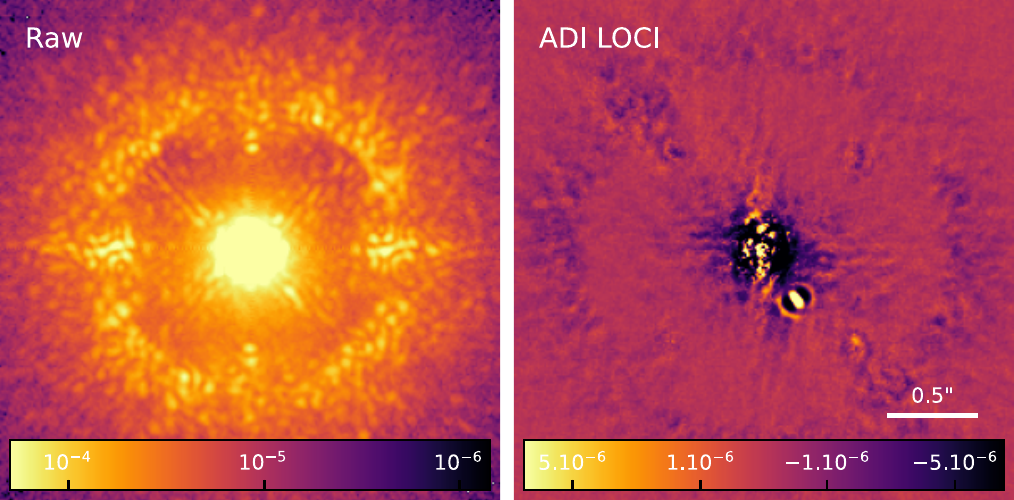}
    \caption{\it {\bf Left}: Raw coronagraphic image recorded by SPHERE/IRDIS with~$2\,$s exposure and dominated by a smooth halo at the center and stellar speckles elsewhere. {\bf Right}: Processed image by~LOCI of~200 raw coronagraphic images recorded with angular differential imaging. The exoplanet $\beta$-Pictoris\,b is located on the bottom-right of the star. Color bars are in normalized intensity.}
    \label{fig:im_diff_beta_pic_sphere}
\end{figure}

To  illustrate this, we show in Figure~\ref{fig:im_diff_beta_pic_sphere} images of the~$\beta-$Pictoris system: raw SPHERE image (left) and after~LOCI processing of one~ADI sequence~(right). Differential imaging attenuates the speckle intensity by a factor of~$\sim50$ in this sequence.
 This enables the detection of a point-like source: planet $\beta$-Pictoris~b. The self-subtraction signature is clearly visible: two azimuthal negative wings around the exoplanet image. Moreover, because speckles are not completely static during the sequence, part of the stellar light is not fully subtracted, especially at small angular separations from the star (center of the image).

\section{Conclusion}
High-contrast imaging is an attractive technique for the analysis of exoplanets and their formation as it provides spectra and astrometry of exoplanets at~$\gtrsim5\,$AU around close by stars, as well as images of young debris and protoplanetary disks, sometimes simultaneously. This allows for the analysis of exoplanetary atmospheres as well as of the interactions between exoplanets and their environment (host stars and circumstellar disks). The method remains very challenging though because of the large luminosity ratio and the small angular separation between the star and its exoplanets. During the last two decades, numerous instrumental techniques have been introduced, especially for coronagraphy, reviewed in this paper.

We first explained why coronagraphs are needed and how they can be designed. We also introduced a formalism to calculate the light distribution in the coronagraphic image, including the presence of wavefront aberrations that create the stellar speckles, greatly limiting the coronagraph performance. We then described how to measure and minimize the stellar speckle intensity using focal plane wavefront sensing and correction. Finally, we presented post-processing techniques using differential imaging and associated observing strategies.

This review showed that there are many techniques for each subsystem of a coronagraph instrument and there is no perfect solution for the various scientific objectives. An instrument optimized for imaging Earth-like planets around Solar-type stars from space will certainly be very different from an instrument optimized for young Jupiter-like planets observed from the ground.

Our field is currently actively involved in the design of the upcoming third generation of exoplanet imagers. For the first time, these instruments are designed from the beginning as fully integrated systems composed of starlight rejection devices~(coronagraphs), wavefront controllers~(adaptive optics and focal plane wavefront control) and post-processing techniques, all working with each others. For ground-based telescopes, the instruments will associate coronagraphy, extreme adaptive optics and focal plane wavefront control: SPHERE+\cite{boccaletti2020_SPHEREImagingYoung} and GPI$2.0$\cite{chilcote2020_GPIUpgradingGemini}. These instruments will probe closer regions to the stars, where multiple exoplanets are expected to be. They will also be a milestone for the conception of exoplanet imaging instruments for the coming~$30\,$m optical telescopes: Planetary Camera and Spectrograph for the Extremely Large Telescope~\cite{kasper2013_RoadmapPCSPlanetary} or Planetary Systems Imager for the Thirty Meter telescope~\cite{fitzgerald2019_PlanetarySystemsImager}. In space, the Coronagraph Instrument of the Nancy Grace Roman Space Telescope will be the first flying instrument to include a focal plane wavefront control using two deformable mirrors~\cite{krist2018_WFIRSTCoronagraphFlight}. This technological demonstrator will be a milestone towards the large space telescopes of the 2050s.

This review mostly covers the existing techniques of high-contrast imaging. Some of these ideas are yet to be experimentally validated, in optical testbeds and on-sky. There are also certainly other ideas to improve the current instruments~\cite{baudoz2017_FutureExoplanetResearch} and any newcomer in the field is more than welcomed.

\section*{Acknowledgements}
\textit{The authors wish to thank Charles Goulas, Axel Potier and Faustine Cantalloube for kindly sharing raw or unpublished data, used in the figures. The authors are immensely grateful to Christian Wilkinson for patiently proof-reading this review.}

\appendix

\section{Notations and acronyms}
\setcounter{table}{0}
\counterwithin{table}{section}
\begin{table}[!ht]
\begin{center}
\begin{tabular}{c|c}
\textbf{Variable}&\textbf{Description}\\
\hline
$t$&Time variable\\
$\omega$&Wave pulsation\\
$\lambda$&Wavelength of observation\\
$\Delta\lambda$&Bandwidth of the spectral filter\\
$\vec{k}$&Wave vector with $\left\|\vec{k}\right\|=k=2\,\pi/\lambda$\\
$p$&Polarization state\\
$\vec{\xi}$&Coordinates in the pupil planes\\
$\vec{x}$&Coordinates in the focal planes\\
$\xi_h$&Coordinates in the pupil planes in the horizontal direction\\
$x_h$&Coordinates in the focal planes in the horizontal direction\\
$\vec{u}_h$&Horizontal unit vector in the focal planes\\
$z$&Longitudinal position from the pupil plane\\
$z_T$&Talbot length\\
$\mathcal{F}$&Fresnel number\\
$D$&Pupil diameter\\
$T_{exp}$&Integration time used to record the image\\
$F_{\mathrm{Y}}$&Flux coming from object~Y (Y=S for star or P for planet)\\
$\mathrm{PSF}(\vec{x})$&Normalized intensity of the star at position~$\vec{x}$ in the focal image\\
$\mathrm{SNR}$&Signal to noise ratio for the exoplanet detection\\
Y& Off-axis planet (Y=P), on-axis star (Y=S) or incoming wave vector $\vec{k}$ (Y=$\vec{k}$)\\
$\Psi_{\mathrm{X},\mathrm{Y}}$&Electric field in the pupil plane~X (X=A, C or $\mathrm{DM_1}$)  for~Y=P, S or~$\vec{k}$\\
$E_{\mathrm{X},\mathrm{Y}}$&Electric field in the focal plane~X (X=B or D) for~Y=P, S or~$\vec{k}$\\
$\mathcal{C}[\Psi]$&Coronagraph linear operator giving the focal field~$E$ from pupil field~$\Psi$\\
$I_{\mathrm{X},\mathrm{Y}}$&Individual (Y=P, S or $\vec{k}$) or total (no Y) intensity in the focal plane~X (X=B or D)\\
$E_{0,\lambda}$&Square of the energy flux coming from the star at wavelength~$\lambda$\\
$P$& Telescope aperture function\\
$A_\lambda$&Pupil apodization function at wavelength~$\lambda$ \textsuperscript{$\dagger$}\\
$M_\lambda$&Focal plane mask function at wavelength~$\lambda$ \textsuperscript{$\dagger$}\\
$L_\lambda$&Lyot stop function at wavelength~$\lambda$ \textsuperscript{$\dagger$}\\
$f$&Optical focal length\\
up/down&Refers to what happens before (up) or after (down) focal plane~B\\
$\phi_{\mathrm{X}}$&Phase aberrations with X=up or down\\
$\phi\DM$&Phase introduced by a deformable mirror. \\
$a_{\mathrm{X}}$&Amplitude (ie transmission) aberrations with X=up or down\\
$\sigma_{\mathrm{X}}$&Optical path difference corresponding to~$\phi_{\mathrm{X}}$ with X=up or DM\\
$N_{\mathrm{act}}$&Number of actuators across the pupil diameter.\\
$\delta$&2D~Dirac delta function\\
$\FT{\Psi}{\vec{x}}$&Fourier transform of~$\Psi$ calculated at coordinate~$\vec{x}$\\
$\mathrm{FT}^{-1}$&Inverse Fourier Transform\\
$\eta_{\mathrm{S}}$& Coronagraph normalized intensity of the stellar light in the focal plane\\
$\eta_{\mathrm{P}}$&Planetary throughput in the coronagraph focal plane\\
$\mathcal{A}$&Region of the focal plane
\end{tabular}
\caption{\it Definition of all parameters.  \textsuperscript{$\dagger$} can be complex to modify both transmission and phase of the wavefront.}
\label{tab:notations}
\end{center}
\end{table}

\begin{table}[!ht]
\begin{center}
\begin{tabular}{c|c}
{\bf Acronyms} & {\bf Telescopes and instruments}\\
\hline
ACS&Advanced camera for surveys (HST instrument)\\
CGI&Coronagraphic instrument (Roman Instrument)\\
CONICA& Near-infrared imager and spectrograph (VLT instrument)\\
GPI&Gemini planet imager\\
HabEx&Habitable exoplanet observatory\\
HST&Hubble space telescope\\
JWST&James Webb space telescope\\
LUVOIR&Large ultraviolet optical surveyor\\
MagAO&Magellan Telescope adaptive optics\\
MIRI&Mid-infrared instrument (JWST instrument)\\
NAOS& Nasmyth adaptive optics system (VLT instrument)\\
NACO& NAOS – CONICA (VLT instrument)\\
NICI&Near-infrared coronagraphic imager (Gemini South instrument)\\
NICMOS&Near-infrared camera and multi-object spectrometer\\
NIRC2&Near-infrared camera 2 (Keck2 instrument)\\
NirCam&Near-infrared camera (JWST instrument)\\
NIRI&Near-infrared Imager (Gemini North instrument)\\
SCExAO&Subaru coronagraphic extreme adaptive optics\\
SPHERE&Spectro-polarimetric high-contrast exoplanet research (VLT instrument)\\
STIS&Space telescope imaging spectrograph (HST instrument)\\
VLT&Very large telescope\\

\hline
 & {\bf High-contrast instrumentation}\\
\hline
AO&Adaptive optics\\
APLC&Apodized pupil Lyot coronagraph\\
DM&Deformable mirror\\
FPM&Focal plane mask\\
FPWFS&Focal plane wavefront sensor/sensing\\
FQPM&Four-quadrant phase mask\\
IWA&Inner working angle\\
LOWFS&Low-order wavefront sensing\\
LS&Lyot stop\\
OWA&Outer working angle\\
SLM&Spatial light modulator\\
WFC&Wavefront control\\
WFS&Wavefront sensing\\

\hline
 & {\bf Post-processing techniques}\\
\hline
ADI&Angular differential imaging\\
CDI&Coherent differential imaging\\
PDI&Polarization differential imaging\\
RDI&Reference differential imaging\\
\hline
 & {\bf Miscellaneous}\\
\hline
IR&Infrared\\
RV&Radial velocity\\
\end{tabular}
\caption{\it Acronyms.}
\label{tab:acronyms}
\end{center}
\end{table}

\bibliographystyle{crunsrt}
\bibliography{academie_biblio} 

\end{document}